\documentclass[structabstract]{aa}  
\usepackage{graphicx}
\usepackage{natbib} \bibpunct{(}{)}{;}{a}{}{,}
\usepackage{color}
\usepackage{txfonts}
\begin{document}
   \title{Numerical Simulations of Highly Porous Dust Aggregates in the Low-Velocity
	  Collision Regime}

   \subtitle{Implementation and Calibration of a Smooth Particle Hydrodynamics Code}

   \titlerunning{Numerical Simulations of Highly Porous Dust Aggregates}

   \author{R.~J.\ Geretshauser
          \inst{1}
          \and
	  R. Speith
	  \inst{1}
	  \and
	  C.\ G\"uttler
	  \inst{2}
          \and
          M.\ Krause
	  \inst{2}
          \and
          J.\ Blum
          \inst{2}
          }

   \institute{Institut f\"ur Astronomie und Astrophysik,
	      Abteilung Computational Physics,
	      Eberhard Karls Universit\"at T\"ubingen,
	      Auf der Morgenstelle 10, D-72076 T\"ubingen, Germany \\
              \email{ralf.j.geretshauser@uni-tuebingen.de}
         \and
	      Institut f\"ur Geophysik und extraterrestrische Physik,
	      Technische Universit\"at zu Braunschweig,
	      Mendelssohnstr. 3, D-38106 Braunschweig, Germany \\
             }

   \date{Preprint online version: January 11, 2010; accepted by {\it Astronomy \& Astrophysics} December 22, 2009}

% \abstract{}{}{}{}{} 
% 5 {} token are mandatory
 
  \abstract
   {A highly favoured mechanism of planetesimal formation is collisional growth. Single
    dust grains hit each other due to relative velocities caused by gas flows in the
    protoplanetary disc, which they follow. They stick due to van der Waals forces and
    form fluffy aggregates up to centimetre size. The mechanism of further growth is
    unclear since the outcome of aggregate collisions in the relevant velocity and size
    regime cannot be investigated in the laboratory under protoplanetary disc conditions.
    Realistic statistics of the result of dust aggregate collisions
    beyond decimetre size is missing for a deeper understanding of planetary growth.}
   {Joining experimental and numerical efforts we want to calibrate and validate a computer
    program that is capable of a correct simulation of the macroscopic behaviour of highly porous 
    dust aggregates. After testing its numerical limitations thoroughly we will 
    check the program especially for a realistic reproduction of the compaction, bouncing 
    and fragmentation behaviour. This will demonstrate the validity of our code, which will
    finally be utilised to simulate dust aggregate collisions and to close the gap of fragmentation
    statistics in future work.}
   {We adopt the smooth particle hydrodynamics (SPH) numerical scheme with extensions for 
    the simulation of solid bodies and a modified version of the Sirono porosity model.
    Experimentally measured macroscopic material properties of $\rm SiO_2$ dust are implemented.
    By simulating three different setups we calibrate and test for the compressive strength
    relation (compaction experiment) and the bulk modulus (bouncing and fragmentation 
    experiments). Data from experiments and simulations will be compared directly.
   }
   {SPH has already proven to be a suitable tool to simulate collisions at rather high 
    velocities. In this work we demonstrate that its area of application can not only
    be extended to low-velocity experiments and collisions. It can also be used to simulate
    the behaviour of highly porous objects in this velocity regime to a very high accuracy.
    A correct reproduction of density structures in the compaction experiment, of the
    coefficient of restitution in the bouncing experiment and of the fragment mass distribution
    in the fragmentation experiment show the validity and consistency of our code
    for the simulation of the elastic and plastic properties of the simulated dust aggregates.
    The result of this calibration process is an SPH code that can be utilised to
    investigate the collisional outcome of porous dust in the low-velocity regime.
   }
   {}

   \keywords{hydrodynamics - methods: laboratory - methods: numerical - 
      planets and satellites: formation - planetary systems: formation - 
      planetary systems: protoplanetary disks}

   \maketitle
%
%________________________________________________________________

\section{Introduction}

In gaseous circumstellar discs, the potential birthplace of planetary systems, 
dust grains smaller than a micrometre grow to kilometre-sized planetesimals, which
themselves proceed to terrestrial planets and cores of giant planets by 
gravity-driven runaway accretion. Depending on their size the dust grains and
aggregates perform motions in the disc and relative to each other. Brownian motion,
radial drift, vertical settling and turbulent mixing cause mutual collisions
\citep{Weidenschilling_1977, Weidenschilling_Cuzzi_1993}. 

Since real protoplanetary dust particles are not available for experiments 
in the laboratory much of the following work has been carried out with dust
analogues such as $\rm SiO_2$ \citep{BlumWurm2008}. Theoretical models
also refer to microscopic and macroscopic properties of these materials.
Initially the dust grains hit and stick on contact by van der Waals 
forces \citep{Heim_etal_1999}. In this process, which has been investigated
experimentally \citep{Blum_etal_2000, Blum_Wurm_2000, Krause_Blum_2004} and 
numerically \citep{Dominik_Tielens_1997, Paszun_Dominik_2006, PaszunDominik2008,
PaszunDominik2009, Wada_etal_2007, Wada_etal_2008, WadaEtal:2009}, 
they form fluffy aggregates with a high degree of porosity.

Due to restructuring the aggregates gain a higher mass to surface ratio
and reach higher velocities. \citet{Blum_Wurm_2000, BlumWurm2008} and 
\citet{Wada_etal_2008} showed that collisions among them lead to fragmentation
and mass loss. Depending on the model of the protoplanetary disc,
which provides the kinetic collision parameters,
this means that direct growth ends at aggregate sizes
of a few centimetres. However, \citet{Wurm_etal_2005} and 
\citet{Teiser_Wurm_2009}
have demonstrated in laboratory experiments in the centimetre regime and
with low-porosity dust that the projectile can stick partially to
the target at velocities of more than 20~$\rm m\,s^{-1}$. Thus,
collisional growth beyond centimetre size seems to be possible and
the exact outcome of the fragment distribution is crucial for the
understanding of the growth mechanism.

Numerical models that try to combine elaborate protoplanetary disc
physics with the dust coagulation problem 
\citep{Weidenschilling_etal_1997, Dullemond_Dominik_2005, Brauer_etal_2008,
Zsom_Dullemond_2008, Ormel_etal_2007, Ormel_etal_2009} have to make
assumptions about the outcome of collisions between dusty objects
for all sizes and relative velocities. Since data for these are
hardly available, in the most basic versions of these models perfect sticking, in
more elaborate ones power-law fragment distributions from experiments
and observations \citep{Mathis_etal_1977, Davis_Ryan_1990, Blum_Muench_1993} are assumed. 
The results of alike simulations highly depends on the assumed
fragmentation kernel. However, the given experimental references
have been measured only for small aggregate sizes. The influence
of initial parameters such as the rotation of the objects or
their porosity have not been taken into account for the size
regimes beyond centimetre size, although they
might play an important role \citep{Sirono2004, Ormel_etal_2007}.

A new approach to model the growth of protoplanetary dust aggregates 
was recently developed by \citet{GuettlerEtal:2009b} and \citet{ZsomEtal:2009} 
who directly implemented the results of dust aggregation experiments into a Monte-Carlo 
growth model. They found that bouncing of protoplanetary dust aggregates 
plays a major role for their evolution as it is able to inhibit further 
growth and changes their aerodynamic properties. Although their model 
relies on the most comprehensive database of dust aggregate properties 
and their collisional behaviour, they were unable to make direct predictions 
for any arbitrary set of collision parameters and were thus obliged to 
perform extrapolations over orders of magnitude. Some of these extrapolations 
are based on physical models which need to be supported by further experiments 
and where this is not possible by sophisticated numerical models such as
the one presented here.

Sticking, bouncing, and fragmentation are the important collision outcomes
which need to be implemented into a coagulation code and affect the results 
of these models. Thus, it is not only important to correctly implement the 
exact thresholds between these regimes but also details concerning the outcome 
such as the fragment size distribution, the compaction in bouncing collisions, 
and maybe even the shape of the aggregates after they merged in a sticking 
collision. Due to the lack of important input information the necessity 
for a systematic study of all relevant collision parameters arises and will 
be addressed in this work.

Because of restrictions in size and realistic environment parameters
this task cannot be achieved in the laboratory alone. There
has been a lot of work lately on modelling the behaviour of 
dust aggregates on the basis of molecular interactions between
the monomers \citep{Paszun_Dominik_2006, PaszunDominik2008, PaszunDominik2009, 
Wada_etal_2007, Wada_etal_2008, WadaEtal:2009}. 
However, simulating dust aggregates with a
model based on macroscopic material properties such as density,
porosity, bulk and shear moduli and compressive, tensile and
shear strengths remains an open field since these quantities
are rarely available. The advantage of this approach over the
molecular dynamics method, which is computationally limited
to a few ten thousand monomers, is the accessibility of
aggregate sizes beyond the centimetre regime.

Recently, \citet{Jutzi2008a} have implemented a porosity model
into the smooth particle hydrodynamics (SPH) code by 
\citet{Benz1994}. It was calibrated for
pumice material using high-velocity impact experiments
\citep{Jutzi2009b} and utilised to understand the formation
of an asteroid family \citep{Jutzi2009a}. However, pumice is a material whose
strength parameters decrease when it is compacted (crushed).
Additionally, the underlying thermodynamically enhanced
porosity model is designed to describe impacts of some
$\rm km\,s^{-1}$. Thus, it is perfectly suitable to simulate
high-velocity collisions of porous rock-like material.

In contrast, collisions between pre-planetesimals occur
at relative velocities of some tens of $\rm m\,s^{-1}$ and
compressive, shear and tensile strengths are
increasing with increasing density \citep{Blum2004, Blum2006,
Guttler2009}. 

\citet{Schafer2007} have used an SPH code based on the porosity 
model by \citet{Sirono2004} to simulate 
collisions between porous ice in the $\rm m\,s^{-1}$ regime. 
They found that
a suitable choice of relations for the material parameters
can produce sticking, bouncing or fragmentation of the
colliding objects. Therefore, they stressed the importance
of calibrating the material parameters of porous matter with
laboratory measurements. 
Numerical molecular-dynamics simulations are
about to use a sufficient number of monomers to be close
enough to the continuum limit and to provide
the required material parameters 
\citep[e.g.][reproducing experimental results of \citealt{Blum2004}]{PaszunDominik2008}.
They represent an important support to the difficult
experimental determination of these quantities.
In \citet{Guttler2009} we have measured
the compressive strength relation for spherical $\rm SiO_2$
dust aggregates for the static case in the laboratory and have given
a prescription how to apply this to the dynamic case
using a compaction calibration experiment and 2D simulations. 
We have pointed out relevant benchmark features. It was shown
that the code is in principal capable of simulating
not only fragmentation but also bouncing, which cannot
be seen in molecular-dynamics codes so far. 

In this work we present our SPH code with its technical details
(Sect.\ \ref{sec:physical-model}), experimental reference
(Sect.\ \ref{sec:experimental-reference}), and numerical properties 
(Sect.\ \ref{sec:numerical-issues}). On the basis of
the compaction calibration simulation we demonstrate
that the results converge for increasing spatial resolution
and choose a sufficient numerical resolution (Sect.\
\ref{sec:resolution-and-convergence}). We investigate
the differences between 2D and 3D numerical setup 
(Sect.\ \ref{sec:geometrical-difference}) and 
thereby improve some drawbacks in \citet{Guttler2009}.
Artificial viscosity will be presented as stabilising
tool for various problems in the simulations and its
influence on the physical results will be pointed out
(Sect.\ \ref{sec:artificial-viscosity}).

Most prominently we continue the calibration process
started in \citet{Guttler2009} utilising two further
calibration experiments for bouncing
(Sect.\ \ref{sec:bulk-modulus}) and fragmentation
(Sect.\ \ref{sec:fragmentation}).
In the end we possess a collection of material parameters,
which is consistent for all benchmark experiments. Finally, 
the SPH code has gained enough reliability to be used to
enhance our information about the underlying physics of
dust aggregate collisions beyond the centimetre regime. In
future work it will be applied to generate a catalogue
of pre-planetesimal collisions and their outcome regarding
all relevant parameters for planet formation. Jointly with
experiments and coagulation models this can be implemented
into protoplanetary growth simulations \citep{GuettlerEtal:2009b,
ZsomEtal:2009} to enhance their reliability and predictive
power.

%__________________________________________________________________

\section{Physical Model and Numerical Method}
\label{sec:physical-model}

\subsection{Smooth Particle Hydrodynamics}

The numerical Lagrangian particle method smooth particle hydrodynamics
(SPH) was originally introduced by \citet{lucy1977} and
\citet{gingold1977} to model compressible hydrodynamic flows in
astrophysical applications. Later, the method has been extended 
\citep{libersky1990}, and improved extensively 
\citep[e.g.][]{libersky1993,Benz1994,randles1996,libersky1997} 
to simulate elastic and plastic deformations of solid materials. 
A comprehensive description of SPH and its extensions can be found
in \citet{monaghan2005}.

In the SPH scheme, continuous solid objects are discretized into 
interacting mass packages, so called ``particles''. These particles form 
a natural frame of reference for any deformation and fragmentation that
the solid body may undergo. All spatial field quantities of the object are
approximated onto the particle positions $\vec{x}_i$ by a discretized 
convolution with a kernel function $W$. The kernel $W$ depends on particle 
distance $|\vec{x}_j - \vec{x}_i|$ and has compact support, determined by
the smoothing length $h$ . We are using the standard cubic spline kernel 
\citep{monaghan1985} but normalised such that its maximum extension is 
equal to one smoothing length $h$. 

We apply a constant smoothing length. This also allows to model 
fragmentation of solid objects in a simple way. Fragmentation occurs 
when some SPH particles within the body lose contact with their 
adjacent particles. Two fragments are completely separated as soon as
their respective subsets of particles have reached a distance of more than 
$2h$ so that their kernels do not overlap any more.

Time evolution of the SPH particles is computed according 
to the Lagrangian form of the equations of continuum mechanics, while
transferring spatial derivatives by partial integration onto the analytically
given kernel $W$. 

\subsection{Continuum mechanics\label{sec:continuum-mechanics}}

A system of three partial differential equations forms the framework of continuum
mechanics. As commonly known they follow from the constraints of conservation of
mass, momentum and energy. The first, accounting for the conservation of mass, is
called continuity equation
\begin{equation}
\frac{{\rm d}\rho}{{\rm d}t} + \rho \frac{\partial v^\alpha}{\partial x^\alpha} = 0.
\label{eq:continuity}
\end{equation}
Following the usual notation $\rho$ and $v^\alpha$ denote density and velocity, 
respectively. Greek indices run from 1 to $d$, the dimension of the problem. Einstein
summing notation holds throughout the entire paper. 
In contrast to the usual SPH scheme, where the SPH density $\rho_i$ is 
calculated directly from the particle distribution, we solve the continuity
equation according to 
\begin{equation}
\frac{\mathrm{d}\rho_i}{\mathrm{d}t} = -\rho_i\sum_j
\frac{m_j}{\rho_j} (v^\alpha_j - v^\alpha_i) \frac{\partial
W^{ij}(h)}{\partial x^\alpha_i}
\end{equation}
\citep[e.g.][]{randles1996}.
Here the sum runs over all interaction partners $j$ of particle $i$,
$m_j$ is the particle mass of particle $j$, and $W^{ij}$ denotes the kernel 
for the particular interaction. Although this approach is more expensive,
as it requires to solve an additional ordinary differential equation for 
each particle, it is more stable for high density contrasts and it avoids 
artifacts due to smoothing at boundaries and interfaces.

The conservation of momentum is ensured by the second equation
\begin{equation}
\frac{{\rm d} v^\alpha}{{\rm d}t} = 
   \frac{1}{\rho} \frac{\partial \sigma^{\alpha\beta}}{\partial x^\beta}.
\end{equation}
In SPH formulation, the momentum equation reads 
\begin{equation}
\frac{\mathrm{d}v^\alpha_i}{\mathrm{d}t} = \sum_jm_j
\left(
\frac{\sigma^{\alpha\beta}_i}{\rho_i^2} +
\frac{\sigma^{\alpha\beta}_j}{\rho_j^2}
\right)\frac{\partial W^{ij}(h)}{\partial x^\beta_i}.
\end{equation}
Due to the symmetry in the interaction terms, conservation of
momentum is ensured by construction. Additionally we apply
the standard SPH artificial viscosity \citep{monaghan1983}. This
is essential in particular for stability at interfaces with highly 
varying densities. The influence of artificial viscosity on our simulation
results is investigated thoroughly in Sec.~\ref{sec:artificial-viscosity}.

The third equation, the energy equation, is not used in our model. Hence, 
we assume that kinetic energy is mainly converted into deformation energy
and energy dissipated by viscous effects is converted into heat and radiated
away. 
The stress tensor $\sigma$ can be split into a part
representing the pure hydrostatic pressure $p$ and a traceless part for the shear
stresses, the so called deviatoric stress tensor $S^{\alpha\beta}$. Hence,
\begin{equation}
\sigma^{\alpha\beta} = -p\delta^{\alpha\beta} + S^{\alpha\beta}.
\end{equation}
Any deformation of a solid body leads to a development of internal stresses in a specifically
material dependent manner. The relation between deformation and stresses is not taken
into account within the regular equations of fluid dynamics. Therefore, they are insufficient to describe a perfectly elastic body and have to be extended. The missing relations are the constitutive equations, which depend on the strain tensor
\begin{equation}
\epsilon^{\alpha\beta} = \frac{1}{2} \left(
	\frac{\partial {x'}^\alpha}{\partial x^\beta}
	+ \frac{\partial {x'}^\beta}{\partial x^\alpha}
	\right) \, .
\end{equation}
It represents the local deformation of the body. The primed coordinates denote
the positions of the deformed body.

Following Hooke's law a proportional relation between deformation is assumed involving the material dependent shear modulus $\mu = \mu(\rho)$, which depends itself on the density:
\begin{equation}
S^{\alpha\beta} \propto 2\mu \left(
	\epsilon^{\alpha\beta} - \frac{1}{d} \delta^{\alpha\beta} 
        \epsilon^{\gamma\gamma}
	\right).
\label{eq:elastic-deviatoric}
\end{equation}  
However, this is only the constitutive equation for the traceless shear part. For the
hydrostatic part of the stress tensor we adopt a modification of the Murnaghan equation
of state which is part of the \citet{Sirono2004} porosity model:
\begin{equation}
p(\rho) = K(\rho_0')(\rho/\rho_0' - 1),
\label{eq:elastic-pressure}
\end{equation}
where $\rho_0'$ is the so called reference density, the density of the material at zero external stress, and $K(\rho)$ is the bulk modulus.

The density dependence of the bulk $K(\rho)$ and shear $\mu(\rho)$ moduli is modelled
by a power law
\begin{equation}
K(\rho) = 2 \mu (\rho) = K_0 (\rho/\rho_i)^\gamma.
\label{eq:bulk-modulus}
\end{equation}
Although according to \citet{Sirono2004} $\rho_i$ is the initial density of the material 
at the beginning of the simulation, we, in contrast, want to ensure that our dust material 
possesses the same bulk modulus $K(\rho)$ even for simulations with different initial
densities. In this work the dust material has two different
densities at the beginning of the bouncing (Sect.\ \ref{sec:bulk-modulus}) and fragmentation
(Sect.\ \ref{sec:fragmentation}) calibration setup. According to \citet{Sirono2004} the
materials should feature two different $\rho_i$. As a consequence $K(\rho)$, and in
particular $K_0$, depends on the initial setup. Since we want to compare and validate
$K_0$ by using two different setups we have to fix a unique $\rho_i$ for all simulations. We
choose $\rho_i$ such that $K(\rho_i) = K_0$ is the bulk modulus of the generic uncompressed
dust material that is produced by the random ballistic deposition (RBD) method
\citep{Blum2004}.

The time evolution of the pressure is directly given by the time evolution of the density (Eq.\ \ref{eq:continuity}) and since the pressure is a scalar quantity it is intrinsically invariant under rotation. However, in order to gain a frame invariant formulation of the time evolution of the deviatoric stress tensor, i.e. the stress rate, correction terms have to be added. A very common formulation for SPH \citep[see e.g.][]{Benz1994,Schafer2007} is the Jaumann rate form
\begin{equation}\label{eq:stress_evolution}
\frac{{\rm d}S^{\alpha\beta}}{{\rm d} t} =
	2\mu \left(
	\dot{\epsilon}^{\alpha\beta} - \frac{1}{d} \delta^{\alpha\beta} 
        \dot{\epsilon}^{\gamma\gamma}
	\right)
	+ S^{\alpha\gamma}R^{\gamma\beta} + S^{\beta\gamma}R^{\gamma\alpha},
\end{equation}
where $R^{\alpha\beta}$ is the rotation rate tensor, defined by
\begin{equation}\label{eq:rotation_rate_tensor}
R^{\alpha\beta} = \frac{1}{2}
	\left(
	\frac{\partial v^\alpha}{\partial x^\beta} -
	\frac{\partial v^\beta}{\partial x^\alpha}
	\right)
\end{equation}
and $\dot{\epsilon}^{\alpha\beta}$ denotes the strain rate tensor
\begin{equation}\label{eq:strain_rate_tensor}
\dot{\epsilon}^{\alpha\beta} = \frac{1}{2}
	\left(
	\frac{\partial v^\alpha}{\partial x^\beta} +
	\frac{\partial v^\beta}{\partial x^\alpha}
	\right) .
\end{equation}
To determine rotation rate and strain rate tensor and thus the evolution 
of the stress tensor Eq.~\ref{eq:stress_evolution} in SPH representation,
the SPH velocity derivatives $\partial v_i^\alpha/\partial x_i^\beta$ have
to be calculated. The standard SPH formulation, however, does not conserve
angular momentum due to the discretization error by particle disorder, which
leads to a rotational instability and in particular inhibits modelling rigid 
rotation of solid bodies. To avoid this error, we apply a correction tensor
$C^{\gamma \beta}$ \citep{Schafer2007}  according to 
\begin{equation}
\frac{\partial v_i^\alpha}{\partial x_i^\beta} = \sum_j
\frac{m_j}{\rho_j} (v^\alpha_j - v^\alpha_i) \sum_\gamma \frac{\partial
W^{ij}}{\partial x^\gamma_i}C^{\gamma \beta},
\label{eq:ext_sph}
\end{equation}
where the correction tensor $C^{\gamma \beta}$ is the inverse of
\begin{equation}
\sum_j \frac{m_j}{\rho_j} (x^\alpha_i - x^\alpha_j) \frac{\partial
W^{ij}}{\partial x^\gamma_i},
\end{equation}
that is
\begin{equation}
\sum_j \frac{m_j}{\rho_j} (x^\alpha_j - x^\alpha_i) \sum_\gamma
\frac{\partial W^{ij}}{\partial x^\gamma_i} C^{\gamma \beta} = \delta^{\alpha
\beta}.
\end{equation}
This approach leads by construction to first order consistency where the errors 
due to particle disorder cancel out and the conservation of angular momentum 
is ensured. Only this allows that rigid rotation can be simulated correctly. 

Finally, the sound speed of the material is given by
\begin{equation}
c_s(\rho_0') = \sqrt{K(\rho_0')/\rho_0'}.
\label{eq:soundspeed}
\end{equation}
Together with Eq.~\ref{eq:bulk-modulus} this relation shows that the soundspeed is a
strong function of density. This behaviour has been seen in moleculardynamics
simulations by \citet{PaszunDominik2008}, but there is no data available from laboratory
measurements.

Up to this point the set of equations describes a perfectly elastic solid body.
Additionally, the material simulated in this work are $\rm SiO_2$ dust aggregates, which
features a high degree of porosity and, thus, plasticity. The modifications accounting
for these features are described in the next section.

\subsection{Porosity and plasticity}
\label{sec:porosity-plasticity}

Following the \citet{Sirono2004} model, the porosity $\Phi$ is modelled by the density
of the porous material $\rho$ and the constant matrix density $\rho_s$
\begin{equation}
\Phi = 1 - \frac{\rho}{\rho_s} = 1 - \phi,
\end{equation}
In the following we will use the filling factor $\phi = \rho/\rho_s$.

We model plasticity by reducing inner stresses given by $\sigma^{\alpha\beta}$.
For this reason we need constitutive relations describing the behaviour of the
material during plastic deformation. These relations are specific for each material
and have to be determined empirically. Particularly for highly porous materials it is
extremely difficult to acquire them. Therefore, it is an advantageous feature of our
model that it is based on measurements by \citet{Blum2004}, \citet{Blum2006} and
\citet{Guttler2009}.

The main idea of the adopted plasticity model is to reduce inner stress once the material
exceeds a certain plasticity criterion. In the elastic case, described by Eqns. 
\ref{eq:elastic-deviatoric} and \ref{eq:elastic-pressure}, inner stresses grow linearly 
with deformation. Hence, the material returns to its original shape at vanishing external
forces. Reducing inner stresses, i.e.\ deviating from the elastic deformation path,
reduces the internal ability of the material to restore its original shape. Therefore,
by stress reduction deformation becomes permanent, i.e.\ plastic. Following and 
expanding the approach by \citet{Sirono2004}, we treat plasticity for the pure 
hydrostatic pressure $p$ and the deviatoric stress tensor $S^{\alpha\beta}$ separately.

\begin{figure}
  \resizebox{\hsize}{!}
  {\includegraphics[angle=0]{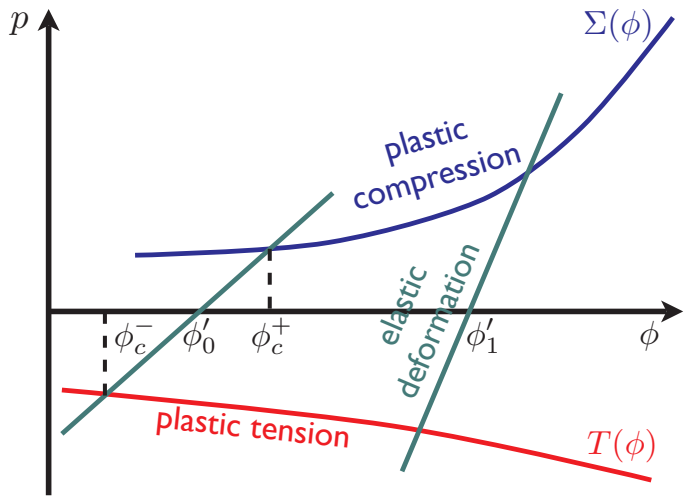}}
    \caption{In the modified Sirono porosity model the regime of elastic deformation
	     is limited by the compressive strength $\Sigma(\phi)$, which represents
	     the transition threshold to plastic compression for $p > \Sigma(\phi)$, 
	     and the tensile strength $T(\phi)$, which represents the transition 
	     threshold to plastic tension or rupture for $p < T(\phi)$. Returning from
	     one of the plastic regimes to vanishing external pressure via an elastic path
	     leads to a $\phi'_1$ that differs from the initial $\phi'_0$. Hence, the
	     material has been deformed irreversibly.}
  \label{fig:sirono-model}
\end{figure}

For the deviatoric stress tensor we follow the approach by \citet{Benz1995} and 
\citet{Schafer2007} assuming that our material is isotropic, which makes the von Mises
yield criterion applicable. This criterion is characterised by the shear strength $Y$,
which in our model is a composite of the compressive and tensile strengths:
$Y(\phi) = \sqrt{\Sigma(\phi) \left| T(\phi) \right|}$. The suitability of this choice
was already demonstrated in \citet{Guttler2009}. Since $Y(\phi)$ is a scalar we
have to derive a scalar quantity from $S^{\alpha\beta}$ for reasons of comparability.
We do this by calculating its second irreducible invariant 
$J_2 = S^{\alpha\beta} S^{\alpha\beta}$. Finally, the reduction of the deviatoric stress
is implemented in the following way
\begin{equation}
S^{\alpha\beta} \rightarrow f S^{\alpha\beta}, 
	\qquad {\rm where} \; f = {\rm min} \left[ Y^2 (\phi) / 3J_2, 1 \right].
\end{equation}
The hydrostatic pressure is limited by the tensile strength $T(\phi)$ for $p < 0$ and
by the compressive strength $\Sigma(\phi)$ for $p > 0$:
\begin{equation}
    p(\phi) = \left\{
    \begin{array}{lc}
    \Sigma(\phi) & \phi > \phi_{\rm c}^+ \\
    T(\phi) & \phi < \phi_{\rm c}^- \\
    \end{array}
    \right.\;.\label{eq:plastic-pressure}
\end{equation}
For $\phi_{\rm c}^- \le \phi \le \phi_{\rm c}^+$ the material is in the elastic regime
and Eq. \ref{eq:elastic-pressure} is applied. $\phi_{\rm c}^-$ and $\phi_{\rm c}^+$ denote
the filling factors where the elastic path intersects the tensile strength and compressive
strength, respectively (see Fig.\ \ref{fig:sirono-model}). 

The pressure reduction process is implemented such that at any time step $p$ is computed
with Eq. \ref{eq:elastic-pressure}. If for a given $\phi$ $p(\phi) > \Sigma(\phi)$ and 
$\phi > \phi_{\rm c}^+$ the pressure $p(\phi)$ is reduced to $\Sigma(\phi)$.
The deformation becomes irreversible once the new reference density
$\rho_0'$ is computed through Eq.\ \ref{eq:elastic-pressure} and the elastic 
path is shifted towards higher densities. Hereby
also the limiting filling factors $\phi_{\rm c}^-$ and $\phi_{\rm c}^+$ are set anew.
In principal there are two possible implementations for this. (1) Plasticity becomes
effective immediately and $\rho_0'$ is computed whenever $p > \Sigma$. (2) Plasticity
becomes effective after pressure decrease, which is equivalent to $\phi < \phi_{\rm c}^+$.
We tested both implementations. For our understanding possibility (1) is closer to
the underlying physical process. In addition it proved to be more stable. According to
the benchmark parameters the results were equivalent.

For the tensile regime, i.e. for $\phi < \phi_{\rm c}^-$, we do not adopt the damage
and damage restoration model presented in \citet{Sirono2004}. This damage model 
for brittle material such as rocks or pumice has been developed
for SPH by \citet{Benz1994, Benz1995} and was recently used by \citet{Jutzi2008a, Jutzi2009a,
Jutzi2009b}. It is assumed that a material contains flaws, which are activated and develop
under tensile loading \citep{Grady1980}. \citet{Schafer2007} did not find the model
applicable in their simulations of porous ice because it includes compressive damage
effects. Brittle material like pumice and rocks tend to disintegrate when they are compressed: they crush.
Whereas in our material, highly porous $\rm SiO_2$ dust, tensile and compressive strength
increase with compression. This is due to the fact that the monomers are able to form
new bonds when they get in contact. Therefore, we adopt the same approach as in the
compressive regime and reduce the pressure $p(\phi)$ to $T(\phi)$ once the
tensile strength is exceeded. Finally, the material can rupture due to plastic flow.
However, material that is plastically stretched can be compressed again up to its full
strength. By choosing this approach a ``damage restoration model'' is implemented
in a very natural way. 

Finally, a remark has to be made about energy. Apart from energy dissipation by numerical
and artificial viscosity we assume intrinsic energy conservation. We suppose that
heat production in the investigated physical processes is negligible. Therefore,
our model is limited to a velocity regime below the sound speed $c_s$ of the dust
material ($\approx$ 30 m/s, see \citealt{BlumWurm2008, PaszunDominik2008}). 
By choosing the approach of modelling plasticity via stress
reduction, we assume that most of the energy is dissipated by plastic deformation, since
the reduction of internal stresses accounts for the reduction of internal energy. 

Since we do not solve the energy equation thermodynamically enhanced features like
any phase transition such as melting and freezing cannot be simulated. This scheme
also does not feature a damage model. Especially in the section about fragmentation
(Sect.\ \ref{sec:fragmentation}) the presence of any flaws in the material cannot
be taken into account yet, although they might have influence on the resulting fragment
distribution.

%\section{Numerical Method}
%\label{sec:numerical-method}
%
%\begin{itemize}
%\item short general introduction into SPH
%\item implementation issues (integration of density vs.\ computation from kernel,
%	implementation of momentum equation)
%\item tensorial correction, rotational instability
%\item artificial viscosity and its dissipative features
%\item spatial vs.\ numerical resolution
%\end{itemize}

\section{\label{sec:experimental-reference}Experimental reference}
\subsection{\label{sec:material_parameters}Material parameters}
The material used for the calibration experiments are highly-porous dust aggregates as described by \citet{Blum2004}, consisting of spherical
 SiO$_2$ spheres with a diameter of 1.5~$\mu$m. For these well defined dust aggregates it is possible to reproducibly measure macroscopic material
parameters like tensile strength, compressive strength, and, potentially, also the shear strength as needed for the SPH porosity model
(see section \ref{sec:porosity-plasticity}). The tensile strength for this material was measured for highly porous and compacted aggregates
($\phi=0.15\;..\;0.66$) by \citet{Blum2004}. These measurements support a linear dependence between the tensile strength and the number of
contacts per monomer (increasing with increasing $\phi$), which yields the tensile strength as
\begin{equation}
    T(\phi) = - \left( 10^{2.8 + 1.48 \phi}\right)\;{\rm Pa} \;. \label{eq:tensile_strength}
\end{equation}
The compressive strength was measured in the experimental counterpart of this paper \citep{Guttler2009} with an experimental setup to determine
the static omni-directional compression (ODC), whereby the sample is enclosed from all sides and the pressure is constant within the sample. We found
that the compressive strength curve can be well described by the analytic function
\begin{equation}
    \Sigma(\phi) = p_{\rm m}\cdot \left(\frac{\phi_2-\phi_1}{\phi_2-\phi}-1\right)^{\Delta\cdot\ln 10}\;,\label{eq:compressive-strength}
\end{equation}
where the free parameters were measured to be $\phi_1=0.12$, $\phi_2=0.58$, $\Delta=0.58$, and $p_\mathrm{m}=13$~kPa. However, these parameters were measured with a static setup and we expect a different strength for a dynamic
collision. The parameters $\phi_1$ and $\phi_2$ determine the range of the volume filling factor, where $\phi_1$ is defined by the uncompressed
material and $\phi_2$ corresponds to the densest packing. These parameters are expected to be the same for the dynamic compression, while $p_\mathrm{m}$ and
$\Delta$ are treated as free parameters, which we will calibrate in the forthcoming sections. 
The last important material parameter to describe the dust aggregates is the elasticity. We can determine the Young's modulus from measurement
and simulation of the sound speed \citep{BlumWurm2008,PaszunDominik2008}, which is $c = 30$~m~s$^{-1}$. From this approach the bulk modulus
for the uncompressed material is $K_0 = \rho_\mathrm{i}c^2 = 300$~kPa with $\rho_\mathrm{i}=300$~kg~m$^{-3}$. However, other plausible
calculations \citep{Weidling2009} indicate an elasticity of rather 1~kPa. We will therefore also vary this parameter.

\subsection{Calibration experiment}
\label{sec:calibration_experiment}

As a setup for an easy and well-defined calibration experiment \citep[see][]{Guttler2009}, we chose a glass bead with a diameter of 1 to 3~mm, which impacts into the dust
aggregate material with a velocity between 0.1 and 1~m~s$^{-1}$ under vacuum conditions (pressure 0.1~mbar). We were able to measure the
deceleration curve, stopping time, and intrusion depth of the glass bead (for various velocities and projectile diameters) and the compaction
of the dust under the glass bead (for a 1.1~mm projectile with a velocity of 0.65~m~s$^{-1}$). These results will serve for calibrating and
testing the SPH code.

For the measurement of the deceleration curve, we used an elongated epoxy projectile instead of the glass bead. The bottom shape and the mass
resembled the glass bead, while the lower density and the therefore longer extension made it possible to observe the projectile during the
intrusion. The projectile was observed by a high-speed camera (12,000 frames per second) and the position of the upper edge was followed with
an accuracy of $\approx 3 \; \mu$m. We found that -- independently of velocity and projectile diameter -- the intrusion curve can be described by
a sine curve
\begin{equation}
    h(t) = - D \cdot \sin \left( \frac{\pi}{2} \frac{t}{T_s} \right)\;; \;\; 
    t \leq T_s.
\label{eq:penetration_curve}
\end{equation}
Here, $D$ and $T_s$ denote the intrusion depth and the stopping time of the projectile. The stopping time is defined as
the time between the first contact and the deepest intrusion at zero velocity. In Sect. \ref{sec:comparison_with_experiments} we will use the normalised form of the
intrusion curve with $D = T_s = 1$.

For the intrusion depth we found good agreement with a linear behaviour of
\begin{equation}
    D = \left(8.3\cdot10^{-4}\;\frac{\rm m^2\;s}{\rm kg}\right) \cdot \frac{mv}{A}\;,\label{eq:penetration_depth}
\end{equation}
where $v$ is the impact velocity of the projectile and $m$ and $A=\pi r^2$ are the mass and cross-sectional area of the projectile,
respectively. We found the stopping time $T_s$ to be independent of the velocity and only depending on the projectile size ($3.0 \pm
0.1$~ms for 1~mm projectiles and $6.2 \pm 0.1$~ms for 3~mm projectiles).

The compaction of dust underneath the impacted glass bead was measured by x-ray micro-tomography. The glass bead diameter and velocity for these
experiments correspond to the compaction calibration setup described in Table \ref{tab:initial-parameters}. The dust sample with embedded glass bead was
positioned onto a rotatable sample carrier between an x-ray source and the detector. During the rotation around $360^{\circ}$, 400 transmission
images were taken, from which we computed a 3D density reconstruction with a spatial resolution of 21~$\mu$m. The according results of the
density reconstruction can be found in \citet{Guttler2009}, where we found that roughly one sphere volume under the glass bead is compressed to
a volume filling factor of $\phi \approx 0.23$, while the surrounding volume is nearly unaffected with an original volume filling factor of
$\phi\approx0.15$. In this work, we will focus on the vertical density profile through the centre of the sphere and the compressed material
(see section \ref{sec:numerical-issues}).

\subsection{Further benchmark experiments}
\label{sec:further-benchmark-experiments}

In this section, we will present two further experiments which will be used for the validation of the SPH code in sections \ref{sec:bulk-modulus} and \ref{sec:fragmentation}.
\citet{HeisselmannEtal:2007} performed low-velocity collisions ($v=0.4$~m~s$^{-1}$) between cubic-shaped, approx. 5~mm-sized aggregates of the
material as described in Sect. \ref{sec:material_parameters} and found bouncing whereby approx. 95\% of the energy was dissipated in a central
collision. Detailed investigation of the compaction in these collisions \citep{Weidling2009} revealed significant compaction of the aggregates
(from $\phi=0.15$ to $\phi=0.37$) after approx. 1,000 collisions. The energy needed for this compaction is consistent with the energy
dissipation as measured by \citet{HeisselmannEtal:2007}. 
\begin{figure}
  \resizebox{\hsize}{!}
  {\includegraphics[angle=90]{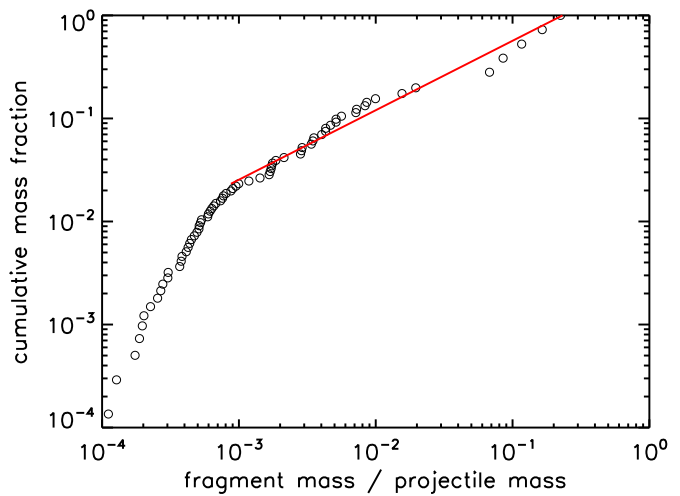}}
    \caption{Cumulative mass distribution of the fragments after a disruptive collision, which can be described by a power law. The divergence for
    low masses is due to depletion of small aggregates because of the camera resolution.}
  \label{fig:frag_size_dist}
\end{figure}

A further experiment deals with the disruptive fragmentation of dust aggregates
\citep[for details see][]{GuettlerEtal:2009b,Guttler2009}. In this case, 
a dust aggregate with a diameter of 0.57~mm consisting of 1.5~$\mu$m spherical 
SiO$_2$ dust with a volume filling factor of $\phi=0.35$ collides with 
a solid glass target at a velocity of 8.4~m~s$^{-1}$ 
\citep[also see Fig. 20 in][]{Guttler2009}. The projectile fragments and the 
projected sizes of these fragments are measured with a high-speed camera with a 
resolution of 16~$\mu$m per pixel.
As the mass measurement is restricted to the 2D images, the projected area of each 
fragment is averaged over a sequence of images where it is clearly separated from 
others. From this projected area, the fragment masses are calculated with the 
assumptions of a spheric shape and an unchanged volume filling factor.
Fig.\ \ref{fig:frag_size_dist} shows the mass distribution in a cumulative 
plot. For the larger masses, which are not depleted due to the finite camera 
resolution, we find good agreement with a power-law distribution
\begin{equation}
    n(m) \; \mathrm{d}m = \left( \frac{m}{\mu} \right)^\alpha \mathrm{d}m \; ,
\label{eq:power-law-distribution}
\end{equation}
where $m$ is the normalised mass (fragment mass divided by projectile mass) and 
$\mu=0.22$ is a measure for the strength of fragmentation, being defined as the 
mass of the largest fragment divided by the mass of the projectile. The exponent 
$\alpha$ has a value of 0.67. A similar distribution was already described by 
\citet{Blum_Muench_1993} for aggregate-aggregate fragmentation of ZrSiO$_4$ 
aggregates with a comparable porosity.

\section{Numerical Issues}
\label{sec:numerical-issues}

Before we perform the calibration process, some numerical issues have to be resolved.
For instance, it is unfeasible to simulate the dust sample, into which the glass bead is dropping
in the compaction calibration experiment presented in section \ref{sec:experimental-reference},
as a whole. It is also unfeasible to carry out all necessary computations in 3D. Therefore,
we will only simulate part of the dust sample and explore at which size spurious boundary
effects will emerge. Most of the calibration process has been conducted in 2D, but the
differences between 2D and 3D results will be discussed and quantified.

In this context we make use of 2D simulations in cartesian coordinates although the
symmetry of the problem suggests the usage of cylindrical coordinates. However, the
SPH scheme in cylindrical or polar coordinates battles with the problem of a singularity
at the origin of the kernel function. There exist only few attempts to solve this
issue \citep[e.g.][]{OmangEtAl:2006}, but they are still under development and require
high implementation efforts. Since in our case 2D simulations only serve as indicator
for calibration and 3D simulations are aimed at, we stick to cartesian coordinates.

\begin{figure}
  \centering
  \resizebox{0.6\hsize}{!}
  {\includegraphics[angle=0]{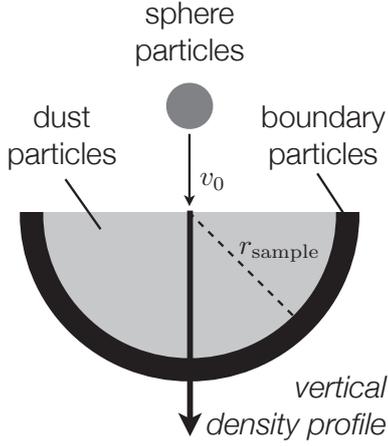}}
    \caption{Compaction calibration setup in 2D or, respectively, cross-section of 3D compaction
	calibration setup. A glass sphere impacts into a dust sample (radius $r_{\rm sample}$) 
	with initial velocity $v_0$.
	The dust sample is surrounded by boundary particles at its bottom. Their acceleration
	is set to zero at every time step. The vertical density profile at maximum intrusion
	serves as calibration feature.}
  \label{fig:standard-setup}
\end{figure}

The glass bead is simulated with the Murnaghan equation of state
\begin{equation}
p(\rho) = \left( \frac{K_0}{n} \right) 
	\left[ \left( \frac{\rho}{\rho_0} \right)^n - 1 \right] \, ,
\end{equation}
following the usual laws of continuum mechanics as presented in section \ref{sec:continuum-mechanics}.
The compaction calibration setup is initialised with the numerical parameters shown in Table
\ref{tab:initial-parameters}, unless stated otherwise in the text. Our tests showed that
the maximum intrusion depth and the density profile are the calibration parameters which are most
sensitive to changes of the numerical setup. Density profiles (e.g. Figs.\ \ref{fig:samplesize-profile}
and \ref{fig:convergence-profile}) display the filling
factor $\phi$ along a line through the centre of the sphere and perpendicular to the
bottom of the dust sample (see Fig.\ \ref{fig:standard-setup}). The origin in the diagrams represents 
the surface of the unprocessed dust sample. The glass sphere has been removed in these figures.

\begin{table}
\centering
\begin{tabular}{l c c l}
\hline\hline
Physical Quantity & Symbol & Value & Unit \\
\hline
\\
\textbf{Glass bead} & & \\
\hline
$\rm {Bulk~density}^{(*)}$ & $\rho_0$	& 2540 			& $\rm kg\,m^{-3}$ \\
$\rm {Bulk~modulus}^{(*)}$ & $K_0$ 	& $\rm 5 \times 10^9$ 	& Pa \\
$\rm {Murnaghan~exponent}^{(*)}$ & $n$		& 4			& - \\
Radius			& $r$		& $0.55 \times 10^{-3}$	& $\rm m$ \\
Impact velocity		& $v_0$		& 0.65			& $\rm m\,s^{-1}$ \\
\\
\textbf{Dust sample} & & \\
\hline
Initial density		& $\rho_i$	& 300			& $\rm kg\,m^{-3}$ \\
Bulk density		& $\rho_s$	& 2000			& $\rm kg\,m^{-3}$ \\
Reference density	& $\rho_0'$	& 300			& $\rm kg\,m^{-3}$ \\
Initial filling factor	& $\phi_i$	& 0.15			& - \\
Bulk modulus		& $K_0$		& $\rm 3 \times 10^5$	& Pa \\
ODC mean pressure	& $p_m$		& 1300			& Pa \\
ODC max.\ filling factor& $\phi_2$	& 0.58			& - \\
ODC min.\ filling factor& $\phi_1$	& 0.12			& - \\
ODC slope		& $\Delta$	& 0.58			& - \\
\hline
\end{tabular}
\caption{Numerical parameters for the compaction calibration setup. ODC stands for
omni-directional compression relation (Eq.\ \ref{eq:compressive-strength}). Quantities
marked by (*) represent the parameters for sandstone in \citet{Melosh1989} which we
adopt for glass here.}
\label{tab:initial-parameters}
\end{table}

\subsection{Computational domain and boundary conditions}

\begin{figure}
  \resizebox{\hsize}{!}
	{\includegraphics[angle=-90, trim=0mm 0mm 12mm 0mm, clip]{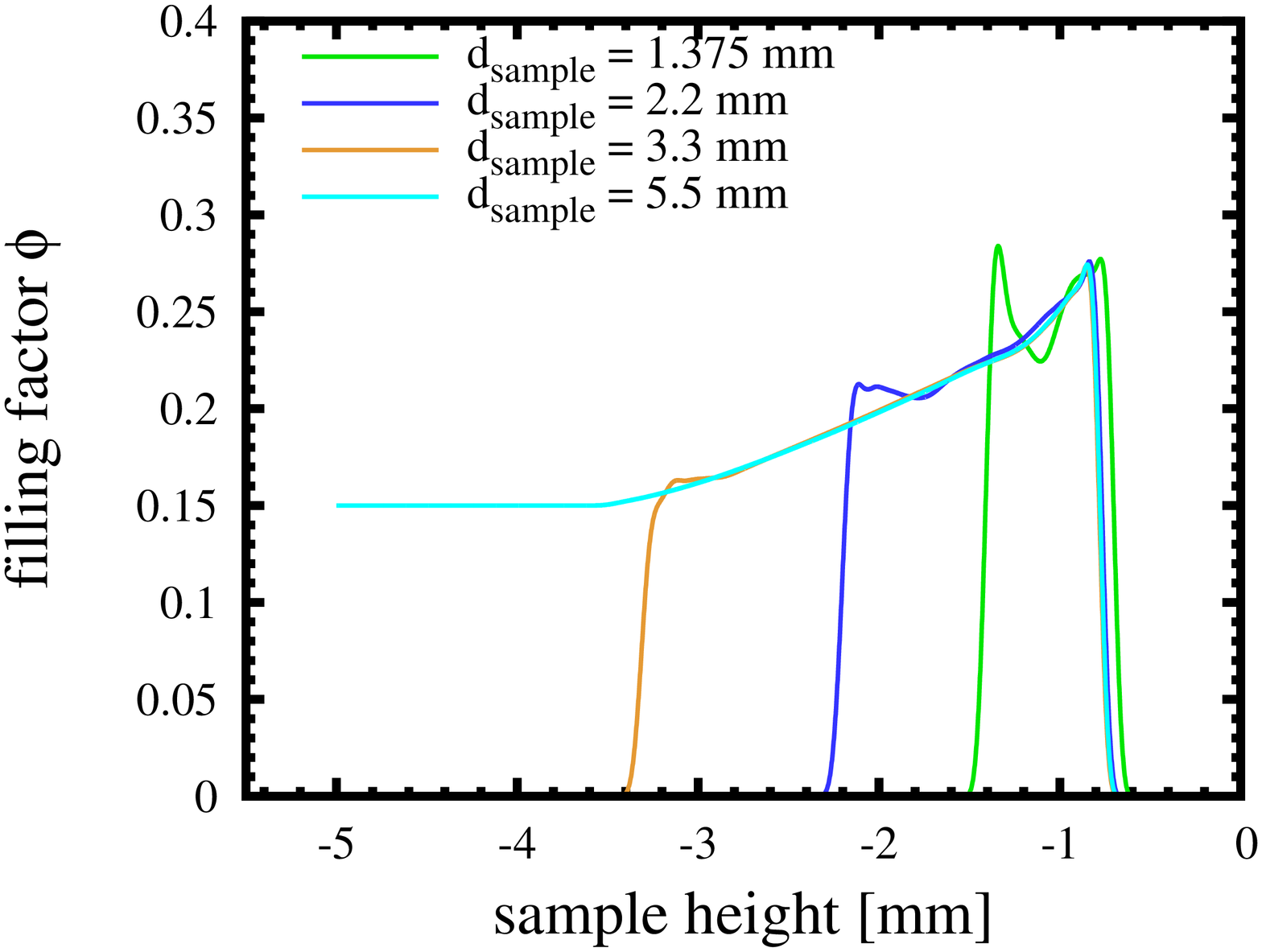}}
  \resizebox{\hsize}{!}{\includegraphics[angle=-90]{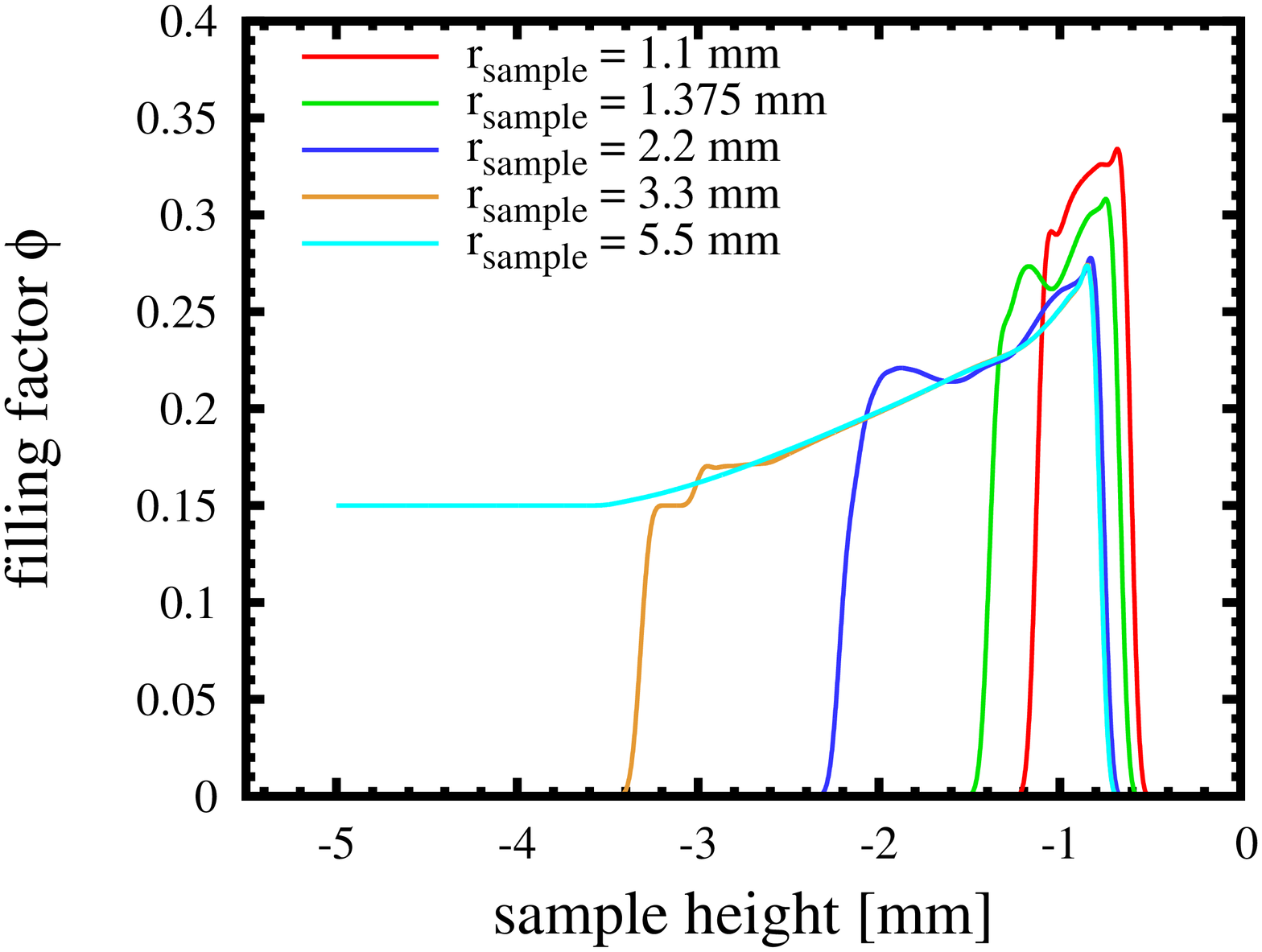}}
  \caption{Vertical density profile at maximum intrusion for the compaction calibration 
	   setup and different shapes of the 2D dust sample (box and semicircle).
	   Depth $\rm d_{sample}$ of an 8 mm wide box (top) and radius $r_{\rm sample}$ 
	   of the semicircle (bottom) were varied. In both cases spurious boundary 
	   effects appear for $d_{\rm sample} < 3.3$~mm and $r_{\rm sample} < 3.3$~mm,
	   respectively.}
  \label{fig:samplesize-profile}
\end{figure}

In 2D simulations we tested the effect of changing size and shape of the dust sample.
Initially the particles were set on a triangular lattice with a lattice constant of
25~$\rm \mu m$. To be geometrically consistent with the cylindric experimental setup,
firstly, we utilised a box (width 8 mm) and varied its depth: 1.375 mm, 2.2 mm,
3.3 mm, and 5.5 mm. This is equivalent to $\rm 2.5 \,\times, 
4 \,\times, 6 \,\times$, and $10 \,\times \,r_{\rm sphere}$. 
Comparing the density profiles (Fig.\ \ref{fig:samplesize-profile}, top), two 
features are remarkable: (1) The maximum filling factor at the top of the dust sample 
($\phi \approx 0.27$ at $D \approx -0.6 \,\rm mm$) and the intrusion depth $D$ is 
nearly the same for all dust sample sizes. (2) For $d_{\rm sample} < 3.3 \,\rm mm$ we 
find spurious density peaks at the lower boundaries ($D \approx - 1.4 \,\rm mm$ and 
$D \approx - 2.2 \,\rm mm$).

In order to reduce the computation time we simulated the dust sample as a semicircle
with the same radius variation as above. The resulting density profiles are shown in
Fig.\ \ref{fig:samplesize-profile} (bottom). In contrast to the corresponding 
simulations with the box-shaped samples we find for $r_{\rm sample} \le 1.375 \,\rm mm$ 
an increased maximum filling factor and a slightly reduced intrusion depth. Due to
the greater amount of volume lateral to the intrusion channel, material can be pushed
aside more easily than inside the narrow boundaries of the semicircle. Therefore, a
higher fraction of the material is compressed to higher filling factors. For
$r_{\rm sample} > 3.3 \,\rm mm$ the spurious boundary effects become negligible within the
compaction calibration setup and the density structure shows no significant difference
for box-shaped and semicircle shaped dust samples.

Hence, all computations of section \ref{sec:numerical-issues} are conducted on the
basis of a semicircle in 2D or a hemisphere in 3D with a radius of $r_{\rm sphere} = 3.3 \,\rm
mm$.
 
In all cases the dust sample is bordered by a few layers of boundary particles. 
The acceleration of these particles is set to zero at each integration step, 
simulating reflecting boundary conditions. Apart from that, i.e. in terms of equation
of state, they are treated like dust particles. We also tested damping boundary
conditions by simulating two layers of boundaries. The outer layer was treated as
described above, the inner (sufficiently large) layer was simulated with a high
artificial $\alpha$-viscosity. Since there was no significant difference in the outcome we fix all boundaries in the afore mentioned way and treat them as reflecting.

\subsection{Resolution and Convergence}
\label{sec:resolution-and-convergence}

\citet{Jutzi2008b} found in his studies of a basalt sphere impacting into a porous target
that the outcome of his simulations strongly depends on the resolution.
With a calibration setup similar to the one used in this paper \citet{Geretshauser2006} confirms that also in simulations of the type presented in this
work a strong resolution dependence is present. He
has found that the intrusion depth of the glass bead can be doubled by doubling the
resolution. Since the calibration experiments presented in \citet{Guttler2009} are
extremely sensitive even to minor changes of the setup, the convergence properties
of porosity model and underlying SPH method will be investigated carefully in this
paragraph. Additionally, we will study the differences in the outcome of 2D and 3D
setup.

For the 2D convergence study particles were initially put on a triangular lattice again. 
The lattice constants $l_c$ were 100, 50, 25, and 12.5 $\rm \mu m$ 
for the compaction calibration setup. 
The smoothing length $h$ was kept constant relative to $l_c$ at a ratio of $5.6 \, \times \, l_c$.
The maximum number of interaction partners was $I_{\rm max} \approx 180$, the average 
$I_{\rm av} \approx 100$ and the minimum $I_{\rm min} \approx 30$.

In the 3D convergence study we are using a cubic lattice with edge lengths $l_c = $ 100, 50,
 and 25 $\rm \mu m$. The latter was simulated with 3.7 million SPH particles, which represent
the limit of our computational resources. We fixed $h = 3.75 \, \times \, l_c$ which
yielded $I_{\rm max} \approx 370$, $I_{\rm av} \approx 240$, and $I_{\rm min} \approx 70$.
\begin{figure}
  \resizebox{\hsize}{!}
	{\includegraphics[angle=-90, trim=0mm 0mm 12mm 0mm, clip]
	                 {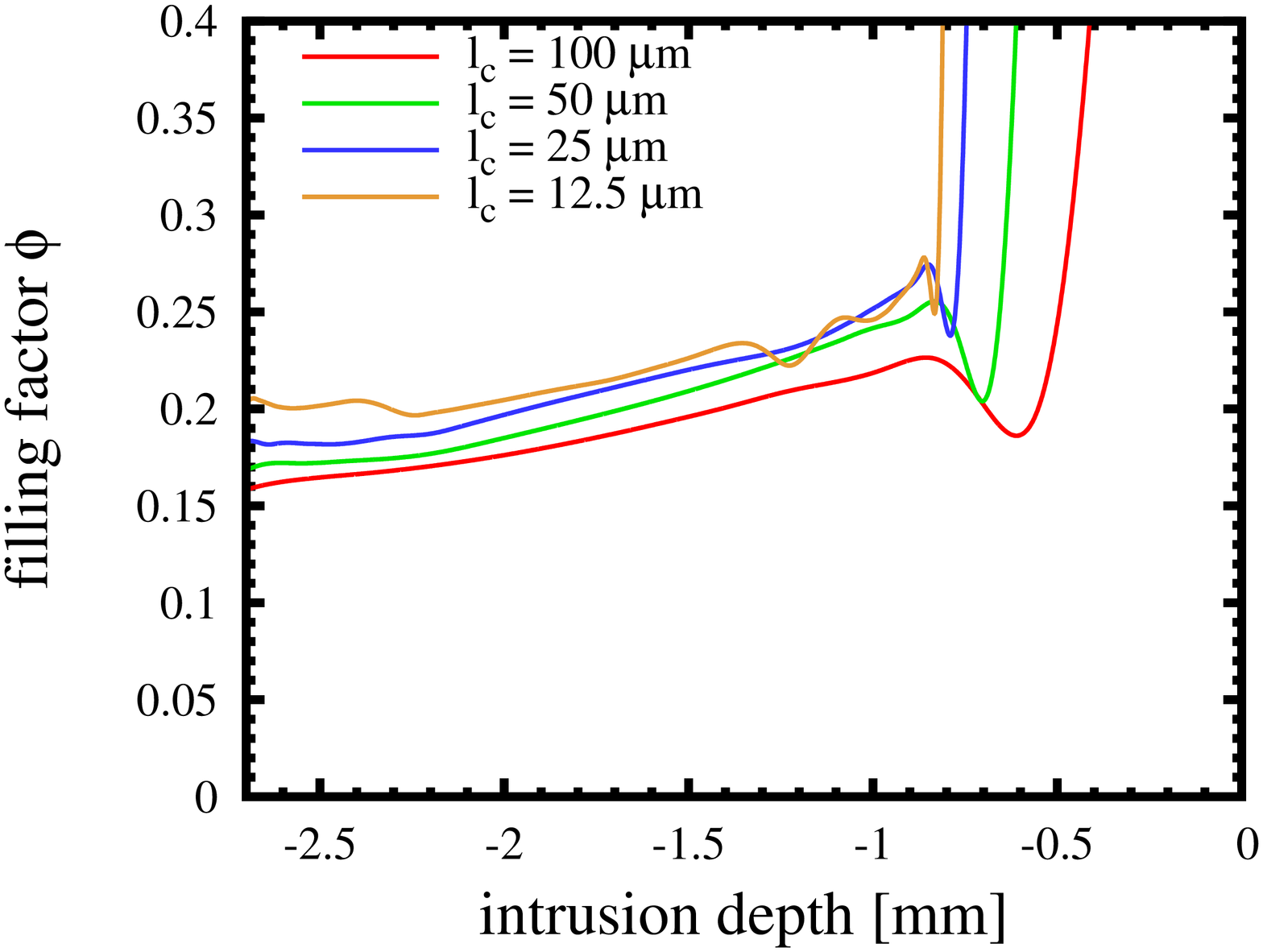}}
  \resizebox{\hsize}{!}{\includegraphics[angle=-90]
			 {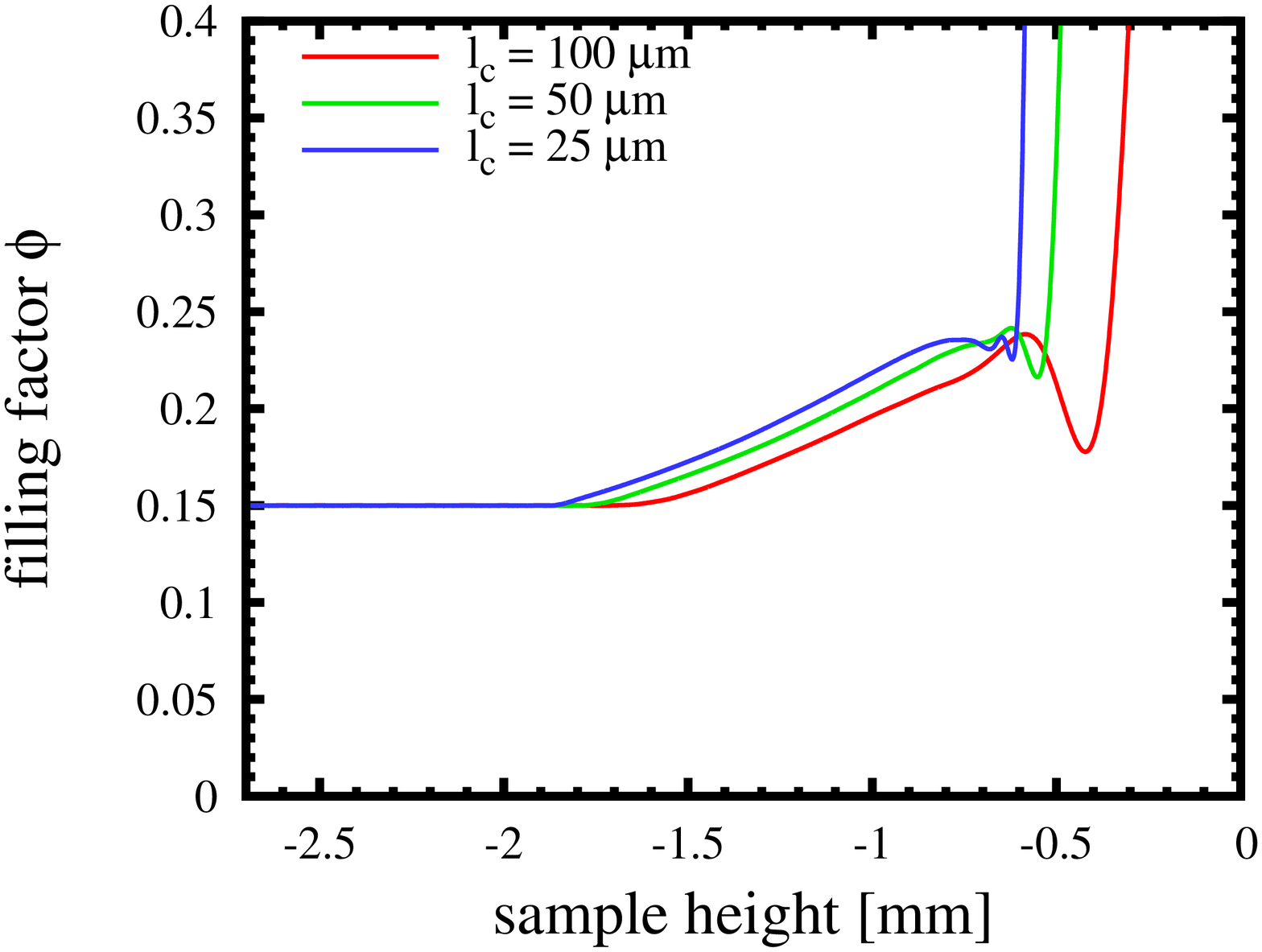}}
  \caption{Convergence study of the vertical density profile for 2D (top) and 
           3D (bottom) compaction calibration setups. 
	   The filling factor
	   increase towards the surface of the dust sample accounts for the glass
	   bead which is not removed in this plot. Simulations were performed for
	   different spatial resolutions. All curves show a characteristic filling
	   factor minimum between sphere and dust sample and a characteristic
	   filling factor maximum indicating the dust sample surface.}
  \label{fig:convergence-profile}
\end{figure}
The results are presented in Fig.\ \ref{fig:convergence-profile}. In contrast to
the plots in Fig.\ \ref{fig:samplesize-profile} the glass sphere is not removed here. Coming
from the right side of the plot, the filling factor rapidly decreases from a 
high value outside the scope of the plot indicating the sphere. The filling factor
reaches its minimum at an artificial gap between sphere and surface of the dust sample.
The width of this gap is about one smoothing length $h$. The existence of the gap
has two reasons: (1) Sphere material and dust material have to be separated by artificial
viscosity due to stability reasons. This will be discussed below. 
(2) The volume of the sphere represents an
area of extremely high density and pressure with respect to the dust sample. This
area is smoothed out due to the SPH method. The width of the smoothing is given by
the smoothing length. 
\begin{figure}
  \resizebox{\hsize}{!}
	{\includegraphics[angle=-90, trim=0mm 0mm 23mm 0mm, clip]
	                 {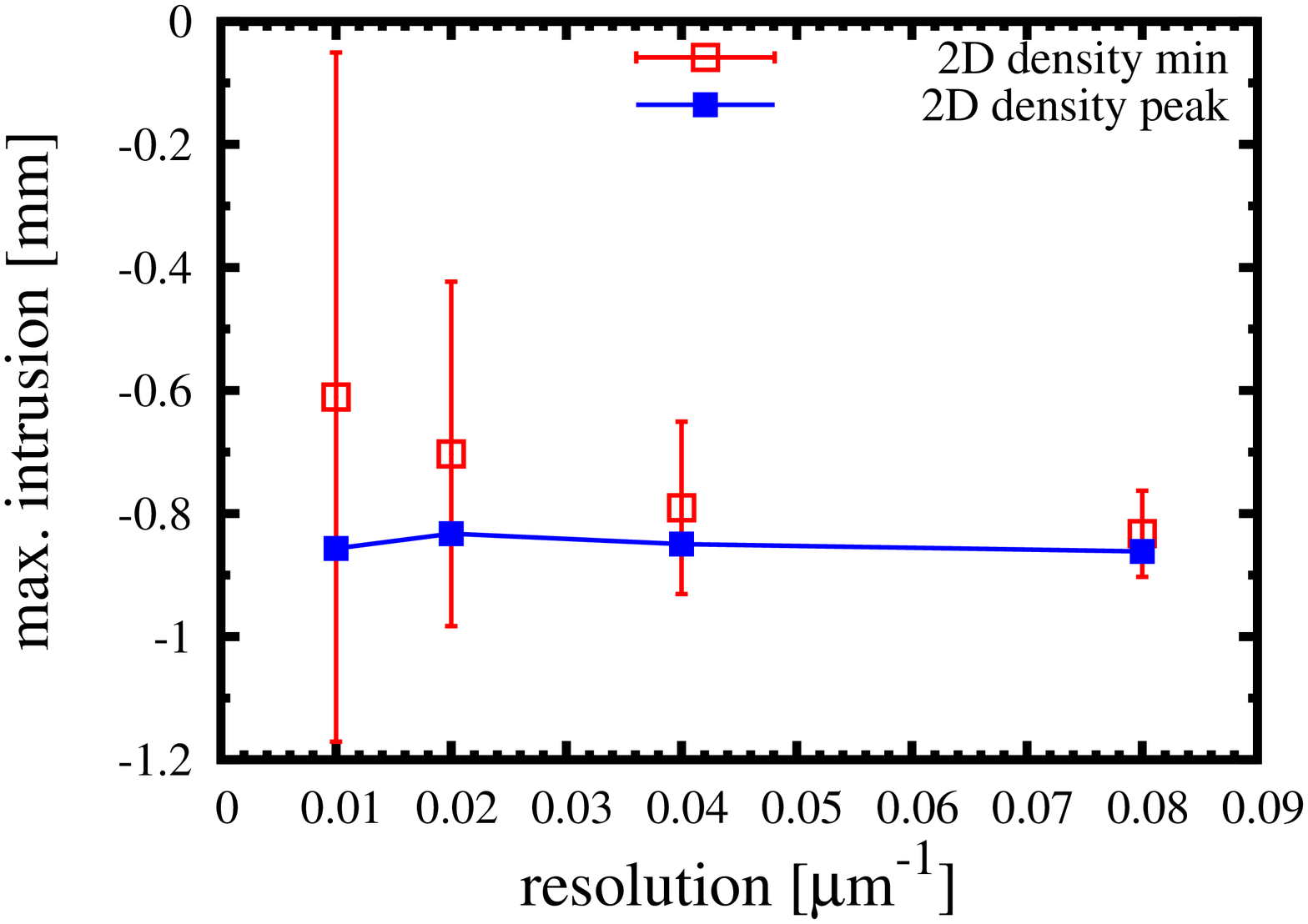}}
  \resizebox{\hsize}{!}{\includegraphics[angle=-90]
			 {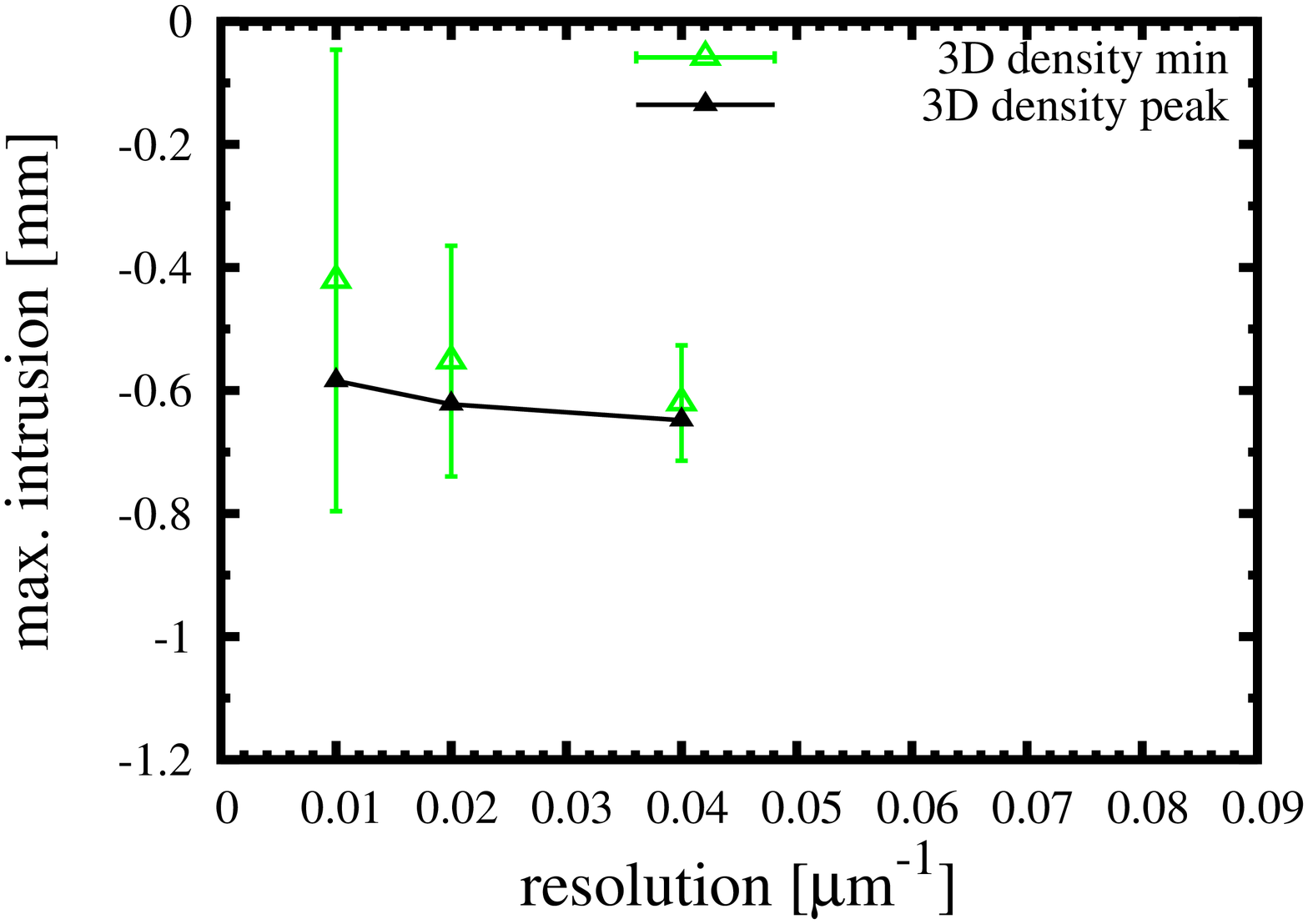}}
  \caption{Convergence study of the maximum intrusion depth for the 2D (top) and
	   3D (bottom) compaction calibration setups.
	   Filled symbols represent the position of the filling factor peak of the
	   dust material, whereas empty symbols denote the position of the minimum
           filling factor at the gap between glass bead and dust material. The values
           are derived from the density profiles in Fig.\ \ref{fig:convergence-profile}.
	   The smoothing length is indicated by the error bars. While the peak position
	   remains almost constant at $\approx -0.9 \,\rm mm$ (2D) and  $\approx -0.65 \,\rm mm$
	   (3D) with increasing spatial 
           resolution, the position of the filling factor minimum quickly converges to
           the same value. This is due to the artificial separation of dust and glass
	   materials, which is in the order of a smoothing length.}
  \label{fig:convergence-intrusion}
\end{figure}
Although a clear convergence behaviour is visible in Fig.\ \ref{fig:convergence-profile}
for both the 2D and the 3D case, a more unique convergence criterion has to be found.
For this purpose we choose the maximum intrusion depth which proved to be very sensitive
to resolution changes. The shape of the filling factor profile offers two choices to
determine the intrusion depth: (1) The filling factor minimum which represents the middle
between sphere and dust sample and (2) the filling factor maximum (peak) of the dust
material left to the gap between sphere and dust sample. 

Fig.\ \ref{fig:convergence-intrusion} shows the results
for both cases in 2D (top) and 3D (bottom). The error bars around the minimum values
represent the smoothing length and give an indication of the maximum error. The
position of the density peak remains almost constant, converging to $\approx -0.9 \,\rm mm$
(2D) and $\approx -0.65 \,\rm mm$ (3D), respectively,
for higher resolutions. The position of the density minimum at low resolutions
significantly differs from the position of the density peak, but converges
quickly to the same intrusion depth with higher resolutions. However, the differences between the extrema
remain well within one smoothing length. This is due to the fact that sphere and
dust sample are separated by about one smoothing length. Comparing 2D and 3D convergence
the 3D case seems to converge more quickly.

Due to the findings of this study we choose a spatial
resolution of $l_c = 25 \,\rm \mu m$ for further simulations in 2D. In the 3D case $l_c = 50 \,\rm \mu m$ is sufficient,
but $l_c \le 50 \,\rm \mu m$ is desirable if it is feasible.
\begin{figure}
  \resizebox{\hsize}{!}{\includegraphics[angle=-90]
			 {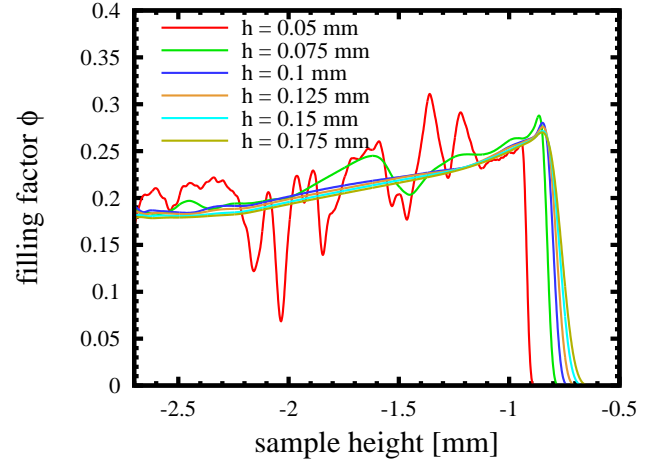}}
  \caption{Convergence study for the density profile using the 2D setup and varying
	   the smoothing length $h$. Through this variation the number of interaction
	   partners is varied according to Table \ref{tab:convergence-wwp}. The glass
	   bead has been removed in this plot. 
	   For $h \le 0.075 \,\rm mm$ clear signs of instabilities are visible. 
	   For $h \ge 0.1 \,\rm mm$ the filling factor has the same value and its
	   position remains constant. The smoothing of the boundary of the dust sample
	   is increased for increasing $h$.}
  \label{fig:convergence-wwp}
\end{figure}
\begin{table}
\caption{Parameters for the convergence study regarding interaction numbers}
\label{tab:convergence-wwp}
\centering
\begin{tabular}{c c c c c c r}
\hline\hline
$h$		& $h/l_c$ & $I_{\rm min}$ & $I_{\rm av}$ & $I_{\rm max}$ & $T_{\rm comp}$ & $N_{\rm steps}$ \\
\hline
\\
0.050 mm	& 2	  & 3 		& 13	    & 25	& 16.2 h     & 132401 	\\
0.075 mm	& 3	  & 10		& 30	    & 55	& 14.8 h     & 96912	\\
0.100 mm	& 4	  & 16		& 53	    & 92	& 19.8 h     & 71175	\\
0.125 mm	& 5	  & 23		& 82	    & 142	& 19.0 h     & 56347	\\
0.150 mm	& 6	  & 32		& 116	    & 205	& 21.6 h     & 46782	\\
0.175 mm	& 7	  & 43		& 158	    & 274	& 24.3 h     & 39980	\\
\hline
\end{tabular}
\end{table}
After defining suitable values for the spatial resolution we now turn to the numerical
resolution, which for the SPH scheme is given by the number of interaction partners of
each single particle. For the investigation of
this feature we performed a study utilising the 2D compaction calibration setup with
a spatial resolution of $l_c = 25 \,\rm \mu m$ and varied the ratio between smoothing
length and lattice constant $h/l_c$ from $2$ to $7$ in steps of one. $h/l_c$ determines
the initial number of interaction partners that is smoothed over. The resulting
maximum, average, and minimum interactions $I_{\rm max}$, $I_{\rm av}$, and $I_{\rm min}$ 
and the corresponding smoothing lengths $h$ can be found in Table \ref{tab:convergence-wwp}.
Additionally, we measured the computation time $T_{\rm comp}$ that the simulations took
on 4 cores of a cluster with Intel Xenon Quad-Core processors (2.66 GHz) for a simulated
time of $5 \,\rm ms$ and the number of integration steps $N_{\rm step}$ of our adaptive
Runge-Kutta Cash-Karp integrator.
 
Comparing the density profiles in Fig.\ \ref{fig:convergence-wwp} (where the glass bead
has been removed) instabilities in the
form of filling factor fluctuations due to insufficient interaction numbers appear
for smoothing lengths $h \le 0.075 \,\rm mm$, i.e.\ for $I_{\rm av} \le 30$. 
For $h \ge 0.1 \,\rm mm$ the density profile maintains essentially the same shape:
The position and height of the filling factor peak remains nearly the same and $\phi$
smoothly drops to $\approx 0.18$ towards the bottom of the dust sample. Only the sharp
edge at the top of the dust sample is smoothed out over a wider range due to the
increased smoothing length.

Table \ref{tab:convergence-wwp} shows that the number of integration steps $N_{\rm step}$
is decreasing with increasing interaction numbers. This is due to the fact that elastic
waves inside the dust sample are smoothed out over a wider range causing the adaptive
integrator to increase the duration of a time step, since density fluctuations do not have
to be resolved as sharply as at smaller smoothing. As expected, the computation 
time $T_{\rm comp}$ is generally increasing with increasing number of interactions. There are
two exceptions: $h = 0.075 \,\rm mm$ and $h = 0.125 \,\rm mm$. Here, the decrease of
$N_{\rm steps}$ overcompensates the increase of interactions leading to a decrease of $T_{\rm comp}$.
Hence, a ratio $h/l_c \approx 5$ yields the necessary accuracy and an acceptable amount
of computation time. This study also justifies the choice of $h/l_c = 5.6$ in
\citet{Guttler2009} and we will stick to this ratio throughout this paper. 

According to these findings, for 3D simulations theoretically an average 
interaction number of $I_{\rm av}^{3/2} \approx 750$ would be needed to achieve 
the same numerical resolution. However, alike simulations are unfeasible and
our choice of $I_{\rm av} \approx 240$ in 3D is equivalent to $I_{\rm av} \approx 40$ in 2D
which should provide sufficient and reliable accuracy.

\subsection{Geometrical difference -- 2D and 3D setups}
\label{sec:geometrical-difference}

\begin{figure}
  \resizebox{\hsize}{!}{\includegraphics[angle=-90]
			 {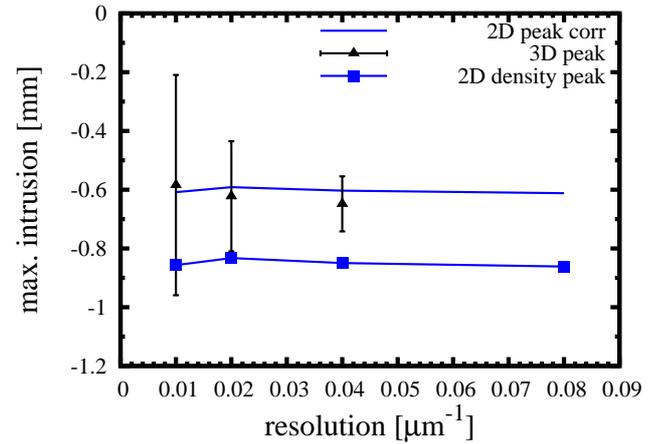}}
  \caption{Verification of the 2D-3D correction factor. Filled symbols denote the position
	   of the filling factor peak of the dust material
	   in Fig.\ \ref{fig:convergence-profile}. Triangles represent 3D and squares 2D
	   values. The conversion from 2D to 3D intrusion depth utilising the correction
	   factor from Eq.\ \ref{eq:correction-factor} and Eq.\ \ref{eq:penetration_depth} 
	   due to the geometrical difference
	   is indicated by the line without symbols. The 3D values are in very good
	   agreement with the very rough theoretical prediction. They lie well within
	   the errors.}
  \label{fig:2D-correction}
\end{figure}

As one can easily see in Fig. \ref{fig:convergence-intrusion} 
2D and 3D simulations have significantly different convergence values
for the intrusion depth. This deviation is caused by the geometrical difference of the
2D and 3D setup. The 2D setup (glass circle impacts into dust semicircle) represents a 
slice through a glass cylinder and a semi-cylindrical dust sample, which
implies an infinite expansion into the third spatial direction. Whereas the 3D setup
represents a real sphere dropping into a ``bowl'' of dust. The relation for the intrusion
depth found by \citet{Guttler2009} (see Sect.\ \ref{sec:experimental-reference}) contains
a geometrical dependence: $D \propto mv/A$, where $D$ is the intrusion depth, $m$ the mass
of the impacting glass bead, $v$ its impact velocity, and $A$ its cross-section.
$m_{\rm 2D}$ is a mass per unit length. \citet{Guttler2009} already exploited 
this relation in order to determine a rough correction factor between 2D and 3D 
simualtional setups:
\begin{equation}
    \frac{m_{\rm 3D}v}{A_{\rm 3D}} = \frac{\frac{4}{3} \pi r^3 \rho \cdot
    v}{\pi r^2} = \frac{8}{3\pi} \, \frac{\pi r^2 \rho \cdot v}{2r}
    = \frac{8}{3\pi} \, \frac{m_{\rm 2D}v}{A_{\rm 2D}} \; .
\label{eq:correction-factor}
\end{equation}
Hence, the 2D intrusion depth has to be corrected by a factor of $\approx \frac{8}{3\pi}$ 
in order to determine the 3D intrusion depth. The comparison is shown in Fig.\ 
\ref{fig:2D-correction}. Again we choose the 2D and 3D data gained in the convergence study
for the peak filling factor values shown in Fig. \ref{fig:convergence-intrusion}.
Fig. \ref{fig:2D-correction} shows the original 2D data, the corrected 2D data,
and the according 3D data (with error bars displaying the smoothing length). The 3D values
nicely follow the rough correction and remain well within the maximum error. This
comparison justifies the correction of the results in \citet{Guttler2009}. With the aid of this
-- now verified -- correction factor the 2D data of the intrusion depth can be converted into
3D data and all calibration tests involving the intrusion depth can be carried out
in 2D, which saves a significant amount of computation time.
Comparing the vertical density profiles of 2D and 3D setups in Fig.\ 
\ref{fig:convergence-profile} resolves also another issue that remained unsolved 
in \citet{Guttler2009}. According to the experimental data the filling factor
drops to a value of $\phi \approx 0.16$ within $\approx 0.6 \,\rm mm$ away from the 
bottom point of
the glass bead. For high-resolution 2D simulations (Fig.\ 
\ref{fig:convergence-profile}, top) the filling factor does not drop to this 
value within the whole dust sample. However, the 3D simulations (Fig.\
\ref{fig:convergence-profile}, bottom) show that this effect again is due to the
difference of 2D and 3D geometry. With the 3D setup the filling factor drops to
$\phi \approx 0.16$ within $\approx 0.9 \,\rm mm$. All deviations from experimental findings
due to this effect, in particular the presence of a large amount of volume with
$\phi \le 0.2$ in the cumulated volume over filling factor diagram 
\citep[Fig.\ 15 in][]{Guttler2009}, can in principal
be removed by switching to 3D simulations. They do not represent a fundamental
error in the porosity model.

\subsection{Artificial Viscosity}
\label{sec:artificial-viscosity}

Since artificial viscosity plays an eminent role for the stability of SPH simulations
we investigate its influence on the outcome
of our compaction calibration setup. Only artificial $\alpha$-viscosity is applied in
our test case, but in a threefold way: (1) It damps high oscillation modes of 
the glass bead caused by the stiff Murnaghan equation of state (EOS). Thereby it 
enlarges the time step of our adaptive integrator and saves computation time.
(2) It is used to provide the dust material
with a basic stability.  (3) It separates the areas of Murnaghan
EOS and dust EOS and prevents a so called ``cannonball instability''. For all three
cases also the influence of $\beta$-artificial viscosity has been tested, but its
influence on all benchmark parameters was negligible.

(1) The choice of the $\alpha$-viscosity of the \emph{glass bead} proved to be very uncritical.
We choose the canonical value $\alpha = 1.0$. There was no influence on the physical
benchmark parameters for all $\alpha$ values, except $\alpha = 0$ which produces an
instability. Values for $\alpha > 1.0$ show no significant effect on the damping and
the influence of $0.1 < \alpha < 1.0$ on it is not too high, but still observable. Hence,
we stick to the canonical value.

\begin{figure}
  \resizebox{\hsize}{!}
	{\includegraphics[angle=-90]
	                 {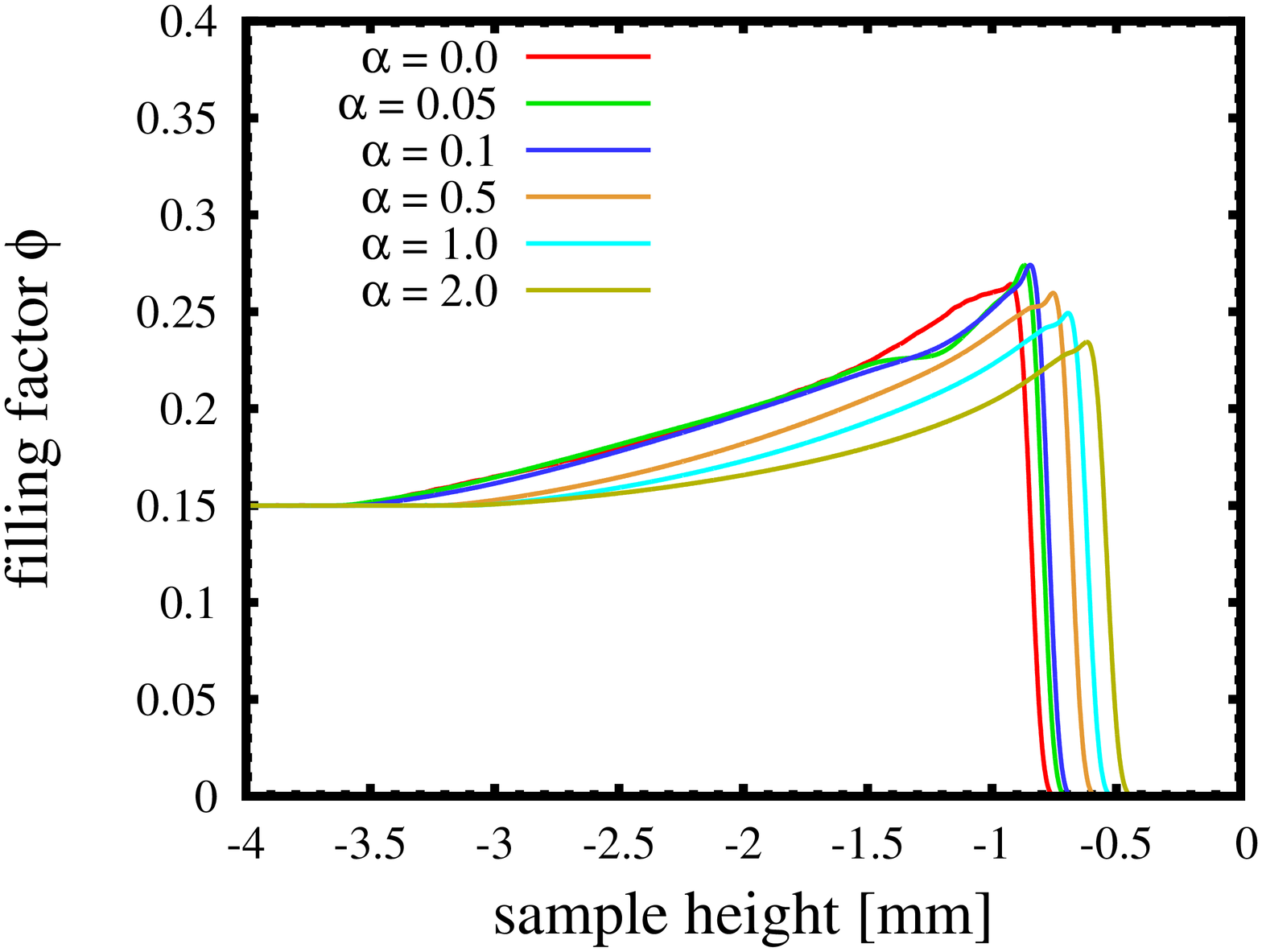}}
  \resizebox{\hsize}{!}{\includegraphics[angle=-90]
			 {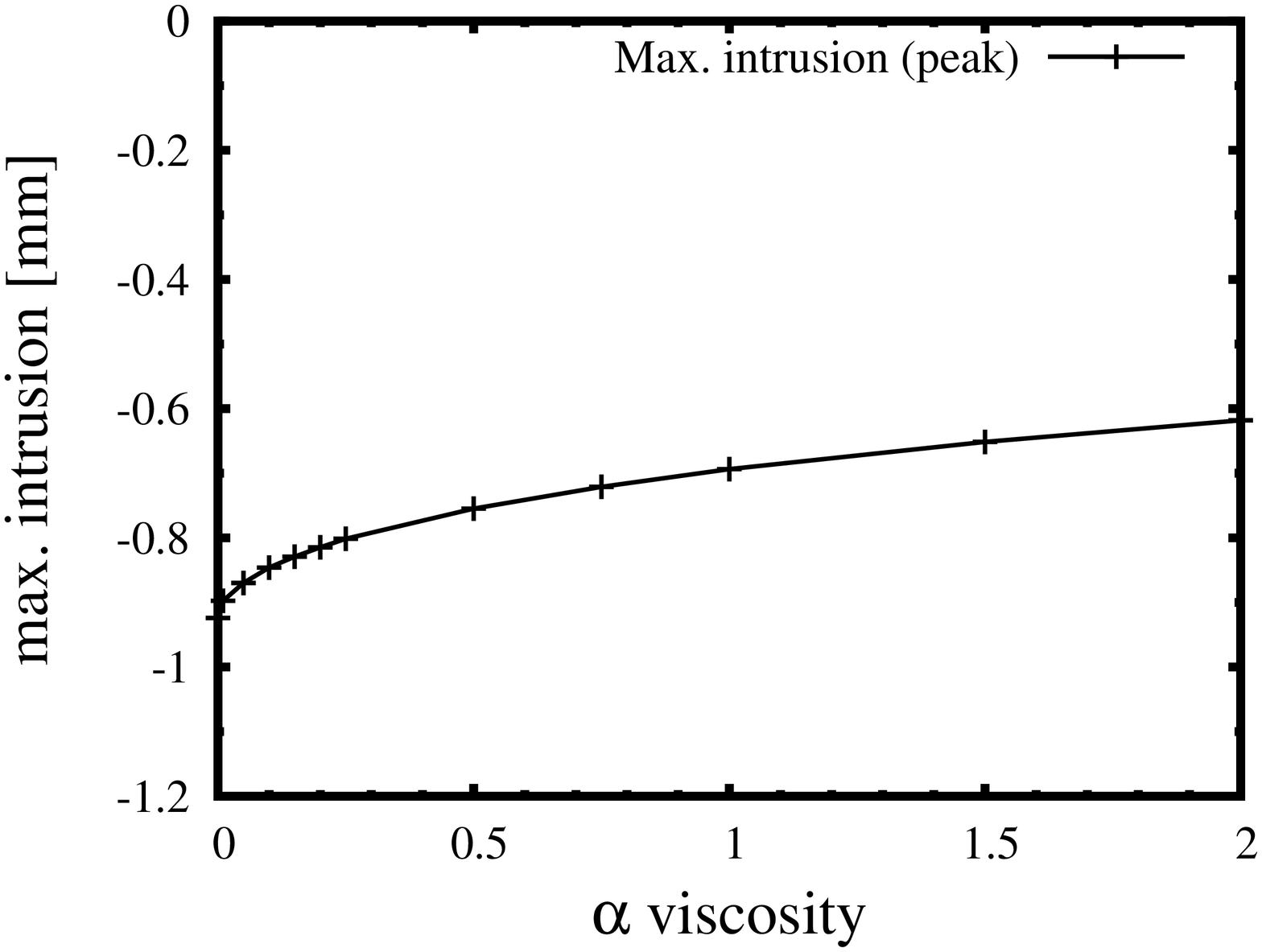}}
  \caption{Density profile (top) and maximum intrusion (bottom) for different 
  values of artificial $\alpha$-viscosity. The shape of the density profile
  hardly changes, but increasing $\alpha$-viscosity decreases the maximum filling
  factor and the maximum intrusion depth.}
  \label{fig:alpha-study}
\end{figure}

(2) \citet{Sirono2004} applies no artificial viscosity to his porous ice material
because of its spurious dissipative properties. Our findings, shown in Fig.\ 
\ref{fig:alpha-study}, confirm this assessment regarding the \emph{dust material}. 
Within our 2D compaction calibration
setup ($l_c = 25 \,\rm \mu m$, $h/l_c = 5.6$) we vary $\alpha$ from $0$ to $2$
and observe its influence on the density profile (Fig.\ \ref{fig:alpha-study}, top)
and the maximum intrusion represented by the filling factor peak of the dust
material (Fig.\ \ref{fig:alpha-study}, bottom).
 
The position of the filling factor peak ranges from $\approx -0.92 \,\rm mm$ with
$\alpha = 0.0$ to $\approx -0.62 \,\rm mm$ at $\alpha = 2.0$. This clearly demonstrates
the dissipative feature of the $\alpha$-viscosity, since a smaller amount of kinetic energy of the
glass bead is transformed into plastic deformation with higher $\alpha$. The
residual energy must have been dissipated. However, the $\alpha$-viscosity-intrusion
curve seems to saturate at a value of $\approx -0.6 \,\rm mm$. The decrease of the maximum
intrusion can also be seen in the density profile (Fig.\ \ref{fig:alpha-study}, top).
While the profile maintains nearly the same shape, the height of the filling factor
peak is decreasing with increasing $\alpha$. Hence, an increasing artificial viscosity
diminishes the peak pressure during compaction, which is via the compressive strength
relation $\Sigma(\phi)$ directly responsible for the height of the filling factor
peak. 

In contrast to \citet{Sirono2004} we find it necessary to apply a small amount
of $\alpha$-viscosity to the dust material. For $\alpha < 0.1$ the results show
traces of an instability, which is also responsible for a rapid increase of the
maximum intrusion. Therefore, we find it convenient to apply an artificial
viscosity with $\alpha = 0.1$ to the dust material, which holds for the previous
simulations of this section as well as the following. The choice of a non-zero $\alpha$, however, is also justified by experimental findings: After impacting into the dust sample the glass bead shortly oscillates due to the elastic properties of the dust. This oscillation is damped by internal friction, which we model with artificial viscosity. Therefore, by choosing a non-zero $\alpha$ we take into
account the dissipative properties that our dust material naturally has. A
quantitative calibration of this parameter, however, has to be left to future work.

(3) During our first simulations with the 2D compaction calibration setup we observed
what is sometimes described in the literature as ``cannonball instability'': During
the compaction process, when the glass bead intrudes into the dust material, single
particles at the sphere's surface start to oscillate between the domains of 
the Murnaghan EOS and dust EOS. Due to the severe discrepancy of the ``stiffness'' 
of these two equations of state the particles collect a huge amount of kinetic energy
until they are fast enough to generate a pressure on the dust material that exceeds
the compressive strength $\Sigma(\phi)$. Eventually they disengage from the
sphere's surface like a cannonball and dig themselves into the dust sample
causing a huge amount of unphysical compaction. We tackle this problem by applying
to all SPH particles with dust EOS, which interact with glass bead SPH particles, the
same amount of $\alpha$-viscosity as to the sphere, i.e. $\alpha = 1.0$. In our
simulations this is sufficient to prevent the ``cannonball instability''.
The spurious dissipation caused by this measure is negligible.

\section{Calibration}

\subsection{Compressive Strength - Compaction Properties}
\label{sec:compressive-strength}

In this section we will refine and extend a study of the compaction properties of the dust 
sample, which was already carried out in a similar, but less detailed way in 
\citet{Guttler2009}. There it turned out that the quantity having the most
influence on the compaction was the compressive strength relation $\Sigma(\phi)$
(Eq. \ref{eq:compressive-strength}), which was measured in an omni-directional
and static manner. In order to adopt this relation for dynamic compression
the mean pressure $p_m$ and the ``slope'' of the Fermi-shaped curve $\Delta$
can be treated as free parameters. The upper and lower boundaries of the
filling factor $\phi_2$ and $\phi_1$, respectively, remain constant even in
the dynamic case. \citet{Guttler2009} found that by lowering $p_m$ most of
the features of the compaction calibration setup can be reproduced in a very
satisfactory manner. Hereby the filling factor over compressive strength 
curve (see also Fig.\ 2 in \citealt{Guttler2009}) is shifted towards lower 
pressures and the yield pressure for compression is lowered.
Using only 2D simulations and a rough parameter grid \citet{Guttler2009} fix 
$p_m = 1.3 \, \rm kPa$. The ``slope'' $\Delta$ has not been considered.
In this work we will consider $\Delta$ and we will perform more accurate
parameter studies for $p_m$. From the latter we will predict a reasonable
choice for $p_m$, which will represent the basis for a 3D simulation of
the compaction calibration setup. The results of this simulation will be
compared to results from the laboratory. For the comparison we use the
same features as \citet{Guttler2009}.

\subsubsection{Fixing free parameters}

\begin{figure}
  \resizebox{\hsize}{!}{\includegraphics[angle=-90]
			 {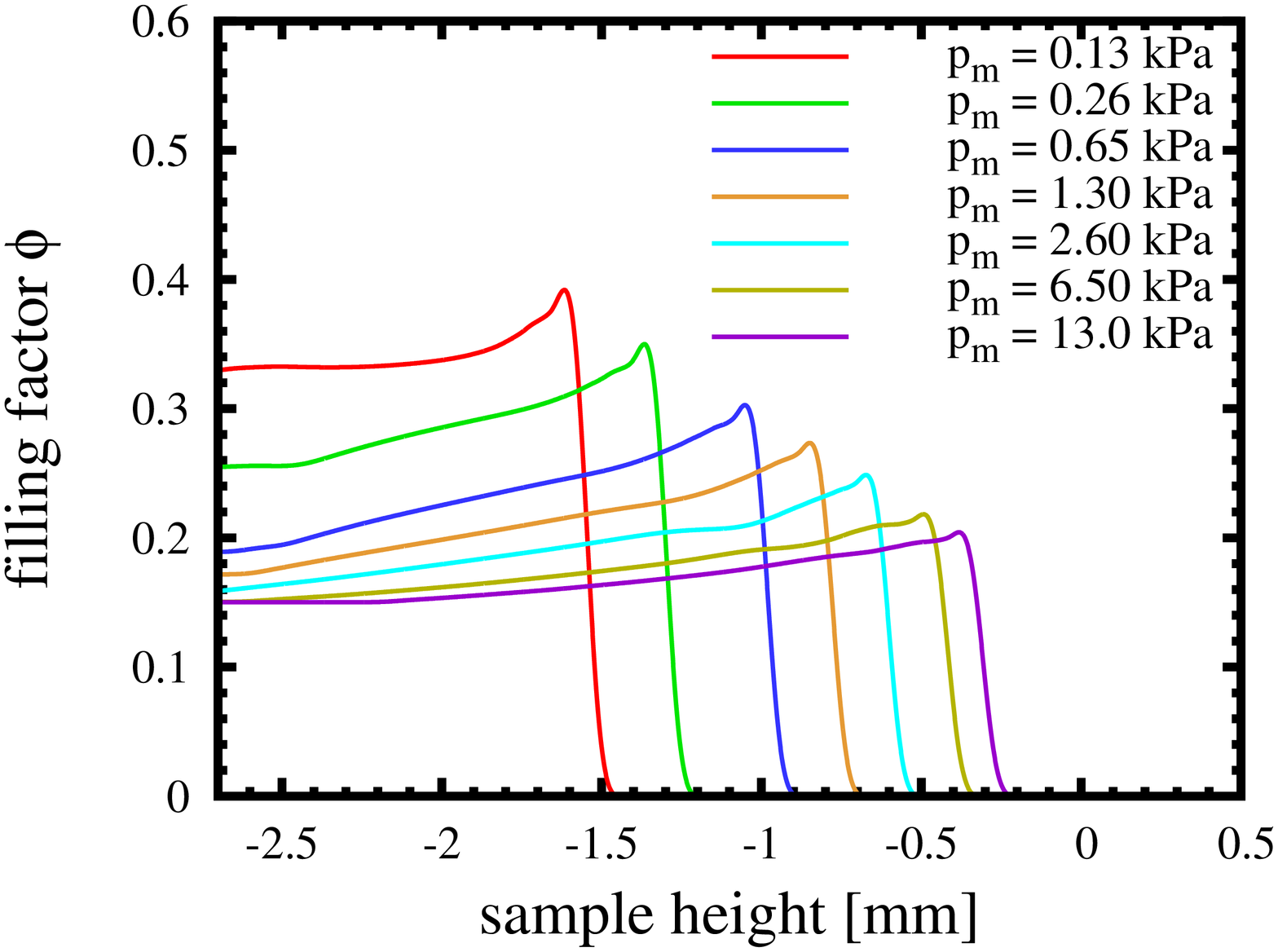}}
  \resizebox{\hsize}{!}
	{\includegraphics[angle=-90]
	                 {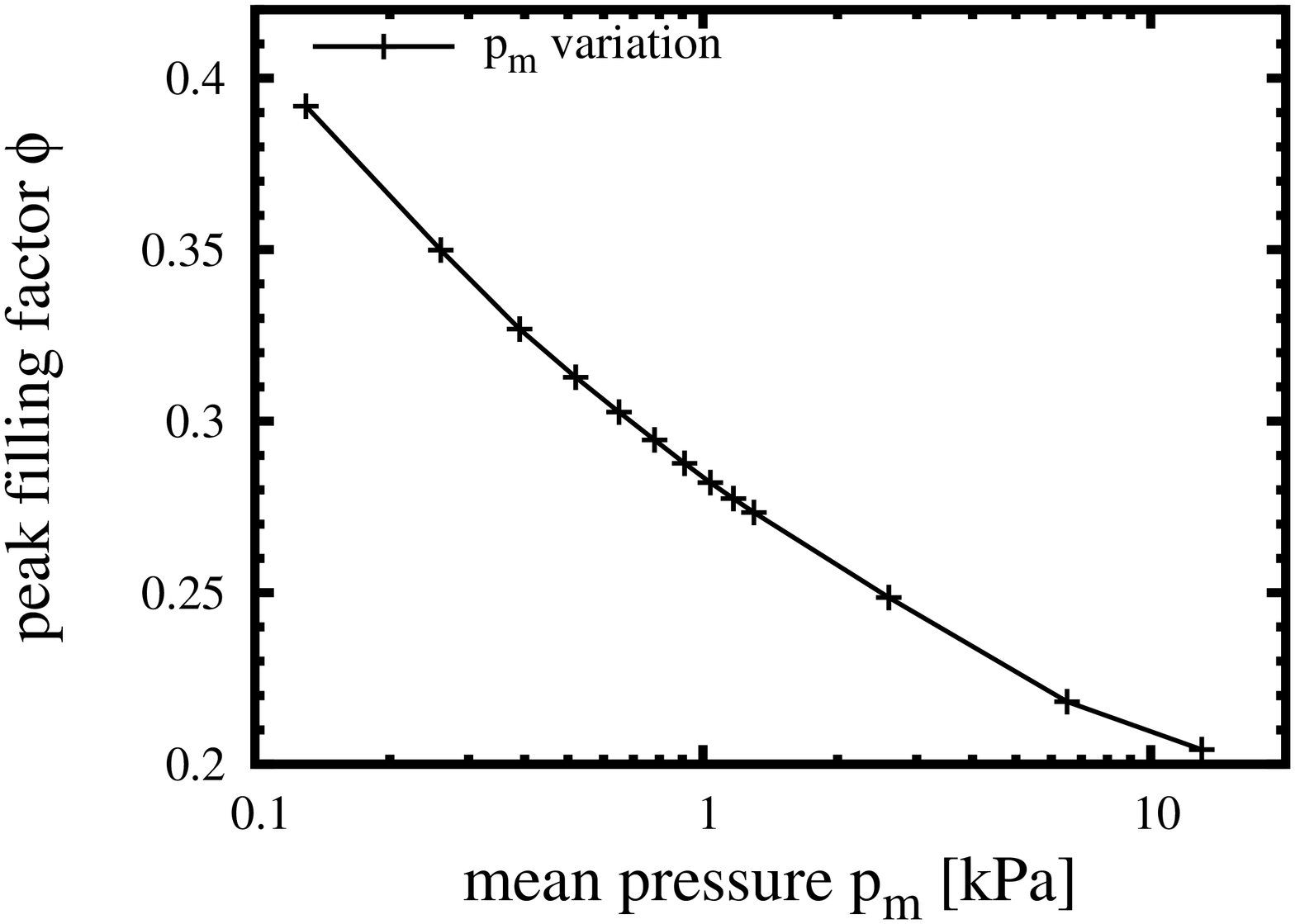}}
  \caption{Density profile (top) and maximum peak filling factor from the density
  		profile over $p_m$ (bottom) for different values of $p_m$, which represents
		the mean pressure in the compressive strength relation $\Sigma(\phi)$
		(Eq. \ref{eq:compressive-strength}). Lowering $p_m$ from
		13.0~kPa (static compressive strength) increases the compressed
		volume, its filling factor, and the maximum intrusion depth of the
		glass bead. The parameter study has been performed using the 2D compaction 
		calibration setup.}
  \label{fig:oacalpha-study}
\end{figure}

\begin{figure}
  \resizebox{\hsize}{!}
	{\includegraphics[angle=-90]
	                 {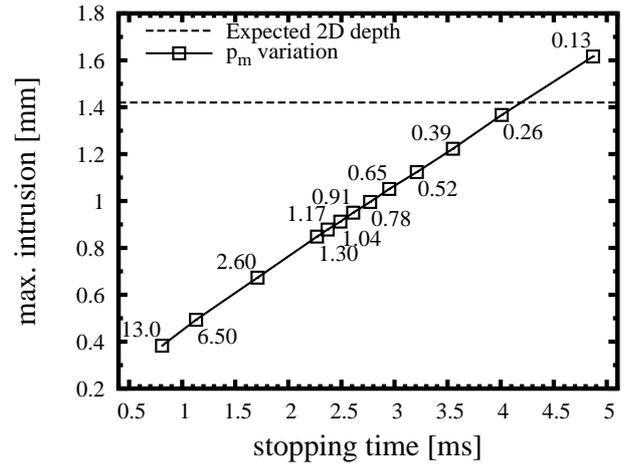}}
  \caption{Maximum intrusion over stopping time for different values of the
  		mean pressure $p_m$ (labels, in kPa) in the compressive strength
		relation $\Sigma(\phi)$ (Eq. \ref{eq:compressive-strength}) using
		the 2D compaction calibration setup. The dashed line indicates
		the 2D intrusion depth that is equivalent to a 3D intrusion depth
	        of $\approx 1$~mm according to Eq.\ \ref{eq:penetration_depth}
	        and the 2D-3D correction factor from Eq. \ref{eq:correction-factor}.
		This supports the choice of the mean pressure $p_m = 0.26$~kPa
	        for further simulations.}
  \label{fig:oacalpha-time-intrusion}
\end{figure}

Since there are no empirical data available for $p_m$ in the dynamical 
compressive strength curve we perform a parameter study in order to 
determine a suitable choice for this important quantity. For this study
we make use of the 2D compaction calibration setup and vary $p_m$ from
$0.13$ to $13.0 \, \rm kPa$ (Fig.\ \ref{fig:oacalpha-study} and 
\ref{fig:oacalpha-time-intrusion}), where $13.0 \, \rm kPa$ represents the value
for the static compressive strength curve. The effect of lowering $p_m$ can most
clearly be seen in the vertical filling factor profile (Fig.\ \ref{fig:oacalpha-study}, top).
First of all, more material can be compressed and it is compressed to higher
filling factors. As a consequence the glass bead intrudes deeper into the
dust sample. From experimental results we expect an intrusion depth of about one
sphere diameter ($\approx 1 \, \rm mm$). With the aid of the empirical
relation between momentum over impactor cross-section $mvA^{-1}$ and
intrusion depth $D$ (Eq.\ \ref{eq:penetration_depth}) as well as the correction
factor between 2D and 3D intrusion depth (Eq.\ \ref{eq:correction-factor}) we estimate that 
$D^{\rm 3D} \approx 1 \, \rm mm$ corresponds to $D^{\rm 2D} \approx 1.42 \, \rm mm$.
Fig.\ \ref{fig:oacalpha-time-intrusion} shows the maximum intrusion over the stopping
time for various values of $p_m$ (labels). The estimated $D^{\rm 2D}$ is
indicated by a dashed line. We deduce that regarding intrusion
depth a dynamic mean pressure $p_m = 0.26$~kPa is a suitable choice.

This is supported by the peak filling factor appearing
in the filling factor profile (Fig.\  \ref{fig:oacalpha-study}, bottom). Empirical
data indicate that in case of the compaction calibration setup a peak filling
factor $\phi_{\rm max} \approx 0.3$ can be expected. The comparison between
2D and 3D results (Sect.\ \ref{sec:geometrical-difference}) has shown
that the peak filling factor in the vertical density profile 
in the 2D case is generally higher than for the same situation in 3D. 
The equivalent of $\phi_{\rm max}^{\rm 3D} \approx 0.3$ is a maximum filling 
factor of $\phi_{\rm max}^{\rm 2D} \approx 0.34$ in 2D. This points to
a choice of $p_m \approx 0.3 \, \rm kPa$, which is consistent with
the findings from the intrusion depth.

\begin{figure}
  \resizebox{\hsize}{!}{\includegraphics[angle=-90]
			 {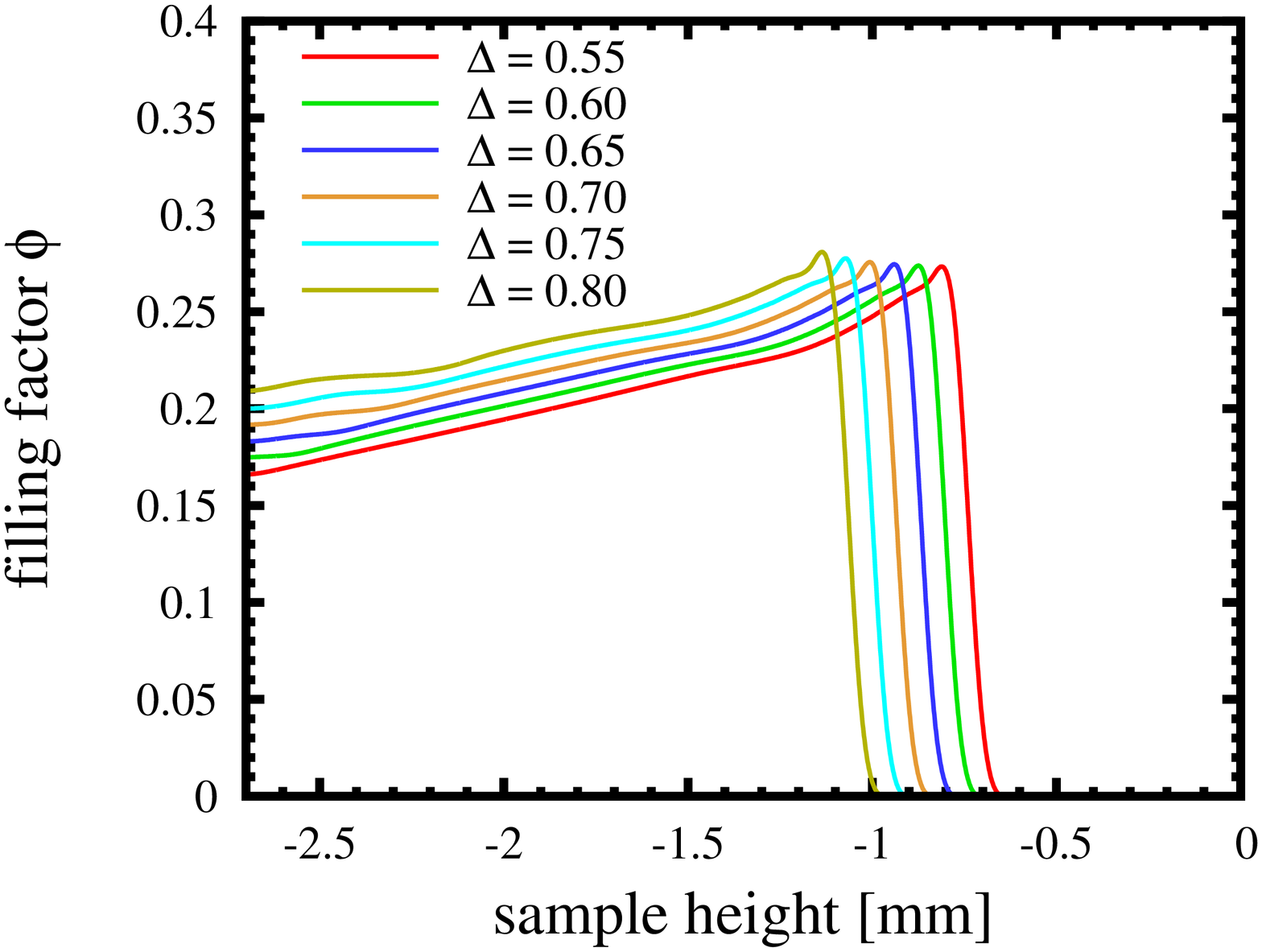}}
  \resizebox{\hsize}{!}
	{\includegraphics[angle=-90]
	                 {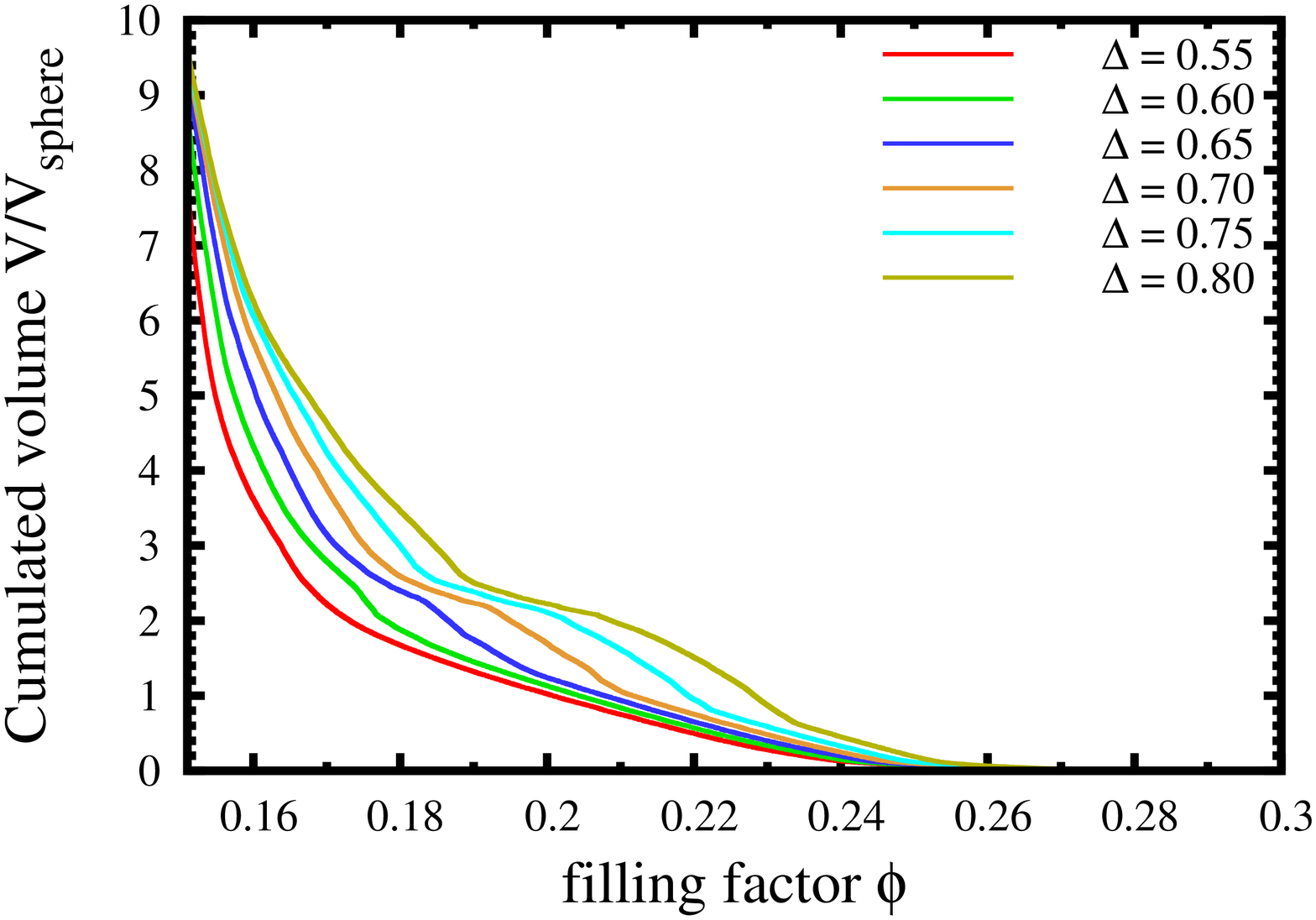}}
  \caption{Density profile (top) and normalised volume fraction of compacted
  		volume over its corresponding filling factor (bottom) for different
		values of $\Delta$ in the compressive strength relation
		(Eq. \ref{eq:compressive-strength}). Increasing $\Delta$ increases
		the amount of compacted volume at filling factors $0.18 < \phi
		< 0.23$ and thereby the maximum intrusion depth. Whereas
		the peak filling factor in the density profile is not changed.}
  \label{fig:delta-study}
\end{figure}

The compressive strength relation $\Sigma(\phi)$ 
(Eq. \ref{eq:compressive-strength}) contains a second free parameter
$\Delta$, which accounts for the ``slope'' of the Fermi-shaped curve.
In \citet{Guttler2009} this parameter was chosen to be the same as the
one of the static omni-directional compressive strength curve and a closer
investigation was not carried out. In order to understand its effect on the
compaction properties of the dust sample we are utilising the 2D compaction 
calibration setup again and vary $\Delta$ from $0.55$ to $0.80$. The
results are presented in Fig.\ \ref{fig:delta-study}. From the vertical density
profile (Fig.\ \ref{fig:delta-study}, top) it can be seen that increasing $\Delta$
increases the intrusion depth, but not as effectively as by lowering the
mean pressure $p_m$. This is due to the fact that increasing $\Delta$ is
hardly increasing the peak filling factor in the vertical density profile.
More volume is compacted to lower filling factors. Whereas by lowering $p_m$
the total amount of volume that is compacted is smaller, but it is compacted
to higher filling factors. This behaviour can be seen comparing the density
profiles for the $p_m$ variation (Fig.\ \ref{fig:oacalpha-study}, top) 
and $\Delta$ variation (Fig.\ \ref{fig:delta-study}, top) and
particularly from the cumulated volume over filling factor
curve (Fig.\ \ref{fig:delta-study}, bottom). This figure shows the normalised
fraction of compacted volume corresponding to a volume filling factor
$> \phi$. The intersection of the curves with the y-axis represents the
total compressed volume, which is increased from $\approx 7$ to
$\approx 9.5$ sphere volumes. According experimental measurements here expect
a value of roughly one sphere volume (see section 
\ref{sec:calibration_experiment}). Especially for $0.18 < \phi < 0.23$ the
compacted volume fraction is increased. The equivalent figure for the $p_m$
variation can be found in \citet[Fig.\ 15]{Guttler2009}.
From the comparison of 2D and 3D
calibration setups (section \ref{sec:geometrical-difference}) we know that
a huge amount of this compaction (especially in the lower filling factor
regime) is due to the geometrical difference. However, the experimental
data do not indicate a particularly high amount of compaction to lower
filling factors (rather the contrary) and, therefore, we maintain our initial
choice of $\Delta = 0.58$.

\subsubsection{Comparison with experiments}
\label{sec:comparison_with_experiments}

For the comparison with experiments we performed a simulation using the 3D
compaction calibration setup (see Table \ref{tab:initial-parameters}) 
with two exceptions: (1) The bulk modulus of the
dust material was set to $K_0 = 2 \, \rm kPa$ (instead of 
$K_0 = 300 \, \rm kPa$) since findings presented below indicated a much lower
bulk modulus. However, the choice of $K_0$ has little influence on the
compaction properties calibrated for in the compaction calibration setup. It is
more important for bouncing and fragmentation, which will be shown below.
(2) The mean pressure of the compressive strength relation 
(Eq.\ \ref{eq:compressive-strength}) was set to $p_m = 0.26 \, \rm kPa$ as
suggested in the previous section.

The following features of the compaction calibration setup were measured in the
laboratory and will be used here for comparison: (1) the stopping time $T_s$,
(2) the intrusion curve of the projectile, (3) the vertical density profile,
(4) the cumulated volume over filling factor curve, and (5) the cross-section
through sphere and dust sample displaying the filling factor.

(1) The experiments show that the stopping time $T_s$ of the glass bead, 
i.e.\ the time from first contact with the dust sample until deepest intrusion, 
is nearly constant at $T_s^{\rm exp} = 3.0 \pm 0.1$~ms for 1~mm projectiles 
over different impact velocities (see section \ref{sec:material_parameters}). In our 
simulation we find $T_s^{\rm sim} = 2.42 \pm 0.05 \, \rm ms$, which is not
in excellent agreement but also not too far off the experimental results. 

\begin{figure}
  \resizebox{\hsize}{!}{\includegraphics[angle=90]
			 {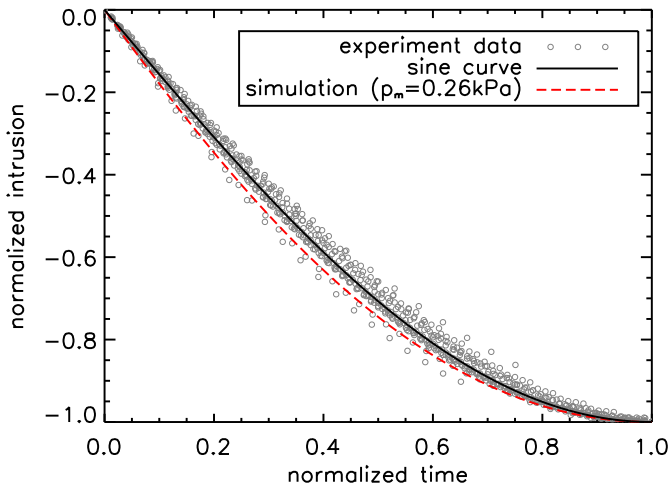}}
  \caption{Comparison between intrusion curves from
	   drop experiments using 1~mm and 3~mm spheres and our
	   3D simulation (mean pressure $p_m = 0.26$~kPa). 
	   The curves are normalised such that the
	   origin represents first touch of sphere and dust sample
	   and (1,-1) denotes stopping time at maximum intrusion.
           The simulated curve lies slightly underneath the fit
	   to the experimental data, but well within the errors.
	   The deviation is due to a smaller stopping time than
	   in the experiments. 
	   }
  \label{fig:norm-3Dfinal}
\end{figure}

(2) The intrusion curve $h(t)$ was cleared from gravity effects and normalised
through $h'(t')=h(t)/D$ and $t' = t/T_S$, where $h(t)$ is the position of the
bottom of the glass bead as a function of time, $D$ the maximum intrusion
depth, and $T_s$ the stopping time. At first contact t is $h'(t'=0) = 0$
and at deepest intrusion $h'(t'=1) = -1$ (see also \citealt{Guttler2009},
section 3.2.2). The comparison is shown in Fig.\ \ref{fig:norm-3Dfinal}: The
intrusion curve generated by our simulation lies well within the data
from the experiments with 1~mm and 3~mm spheres and only slightly below
the sine curve fitted to the empirical data.
 
\begin{figure}
  \resizebox{\hsize}{!}{\includegraphics[angle=-90]
			 {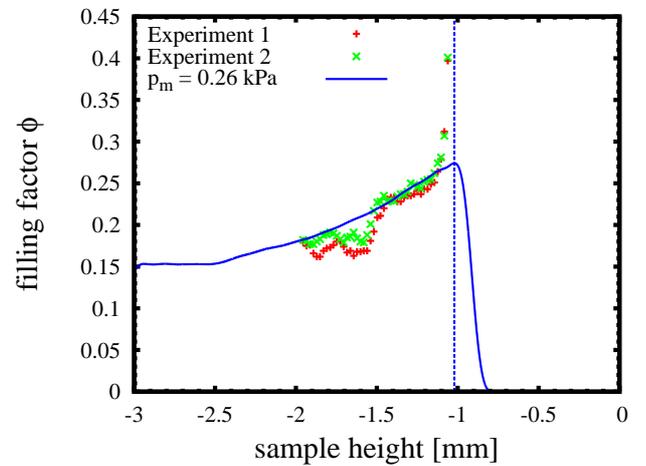}}
  \caption{Comparison between experimentally measured (crosses, sphere
	   not removed) and simulated (solid line, sphere removed) 
	   density profiles at maximum intrusion for the compaction
	   calibration setup. The dashed line indicates the position of
	   the simulated maximum intrusion depth given by the density
	   peak at $-1.02$~mm. The simulation was carried out in 3D
	   using a mean pressure $p_m = 0.26$~kPa for the compressive
	   strength relation (Eq. \ref{eq:compressive-strength}). Both
	   profiles are in excellent agreement. The fact that the step-like
	   structure of the experimental data cannot be seen in the
	   simulation is a minor drawback since it is interpolated nicely.}
  \label{fig:vert-3Dfinal}
\end{figure}

(3) The comparison between the experimentally measured vertical density
profile and the result from our simulation is shown in Fig.\
\ref{fig:vert-3Dfinal}. The crosses represent the data from two
experiments where the sphere has not been removed for the measurement.
For this reason the filling factor reaches extremely high values
at $\approx \, -1$~mm indicating the bottom of the glass bead.
The vertical density profile of our simulation is given by the solid
line and the vertical dashed line is placed at its filling factor
peak at $-1.02$~mm representing the maximum intrusion depth. Comparing
with the experimentally measured maximum intrusion depth of $-1.07$~mm
this is an excellent result. Since a depth of $\approx -1$~mm was
aimed for using Eqns.\ \ref{eq:penetration_depth} and
\ref{eq:correction-factor} and the 2D intrusion depth study of the
previous section this result also supports the validity of these
relations. In addition to the exact value of the intrusion depth
our simulation also reproduces the shape of the given experimental
vertical density profile very well. Only the step-like structure
at $\approx -1.5$~mm does not find an equivalent in our simulation
but it is nicely interpolated.

\begin{figure}
  \resizebox{\hsize}{!}{\includegraphics[angle=-90]
			 {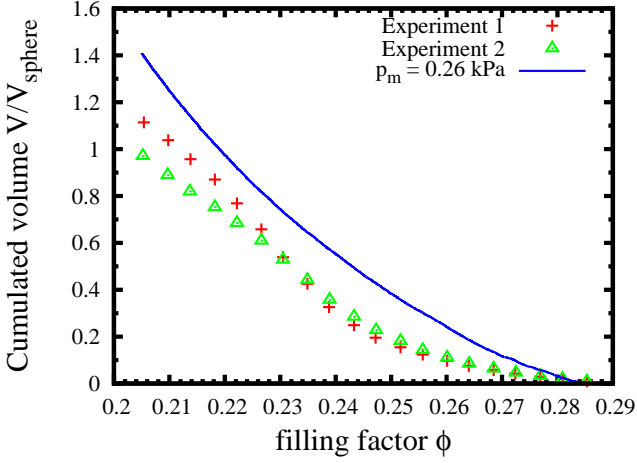}}
  \caption{Total cumulated volume over filling factor for drop experiments
	   (crosses and triangles) and our 3D simulation with mean pressure
	   $p_m = 0.26$~kPa (solid line). The plot shows the cumulated mass
	   fraction (normalised through the sphere volume) with a filling
	   factor $> \phi$ over $\phi$. Our simulation is in good agreement
	   with the experimental findings. However, a higher amount of volume 
	   compressed to high filling factors leads to an almost constant
	   deviation for $\phi < 0.26$. The slope is reproduced very well.}
  \label{fig:mass-3Dfinal}
\end{figure}

(4) The vertical density profile shows only a cut through the
compressed volume. It contains information about the exact structure
of the compression. The cumulated volume over filling factor curve
(Fig. \ref{fig:mass-3Dfinal}) has the advantage that it represents
the total compressed volume together with its filling factors,
hence, these features are not fully independent but focus on different
aspects of the compression. The cumulated volume is normalised
through the sphere volume. It accounts for the volume fraction
of compacted area corresponding to a volume filling factor $> \phi$.
In general, the experimental reference and our simulation show a very
good agreement. Slightly too much volume is compressed to high
filling factors in the simulation which leads to an almost constant
deviation for $\phi < 0.26$. However, the slope is reproduced very
well.

\begin{figure}
  \resizebox{\hsize}{!}{\includegraphics[angle=90]
			 {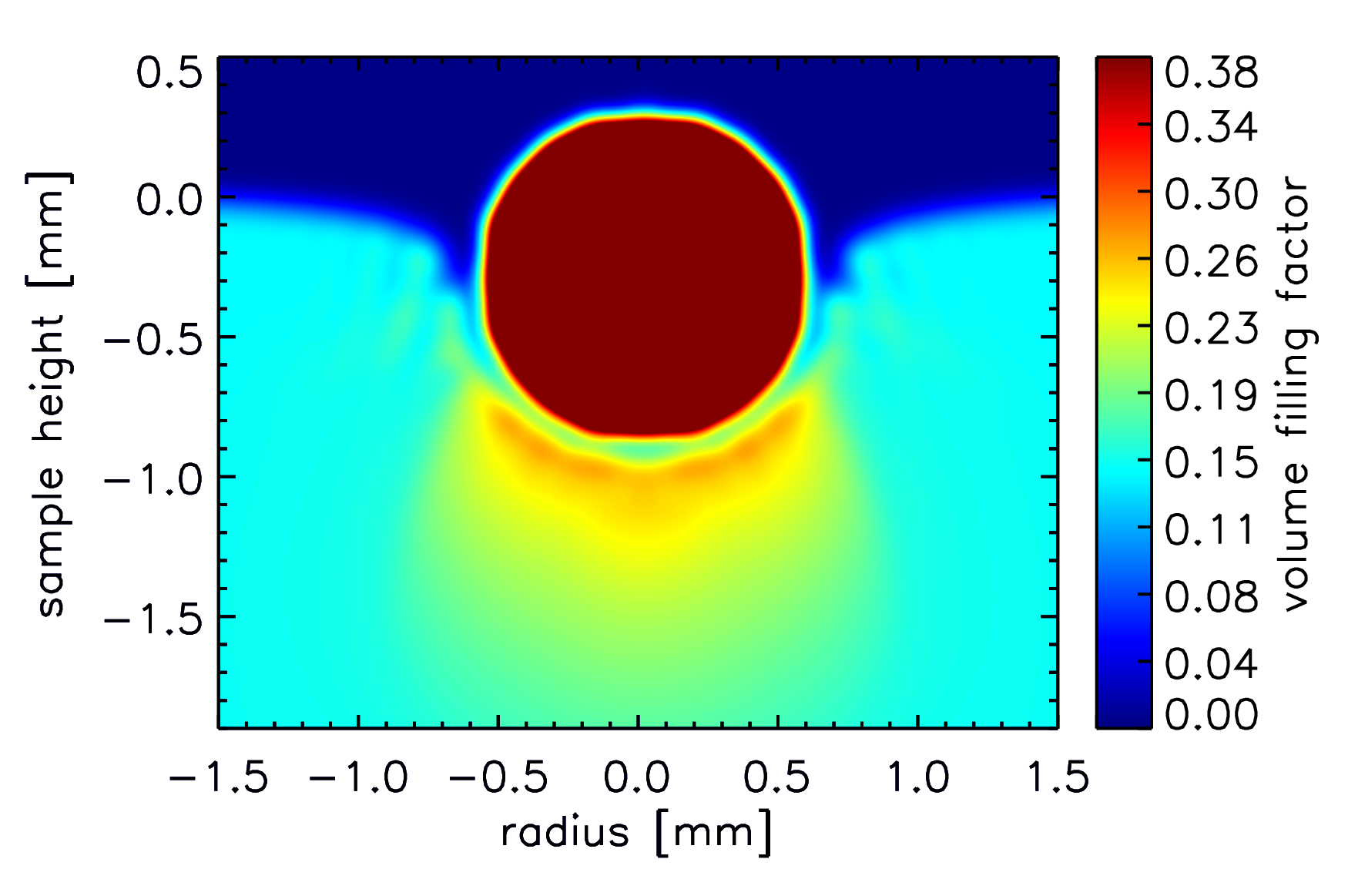}}
  \resizebox{\hsize}{!}{\includegraphics[angle=90]
			 {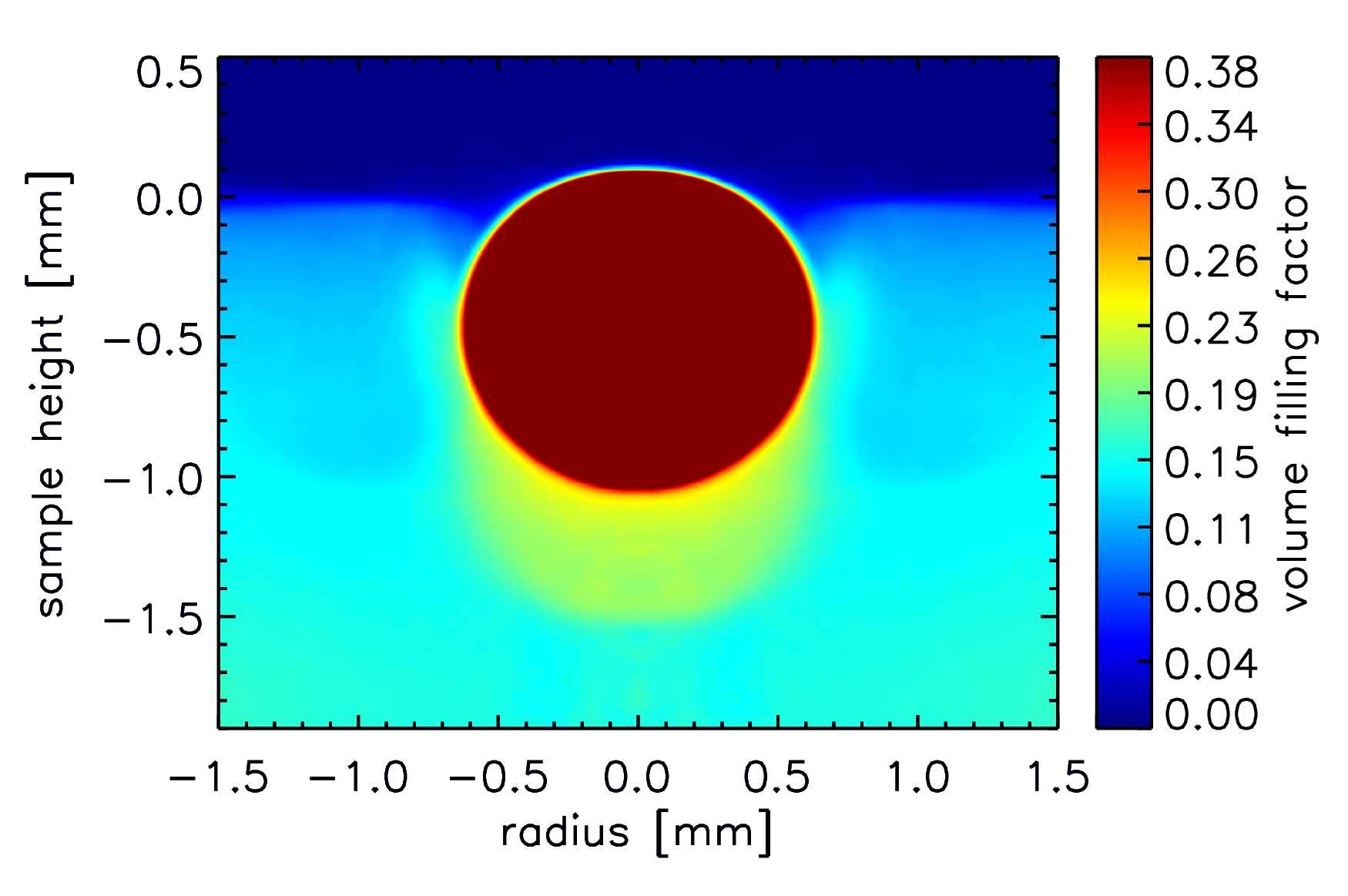}}
  \caption{Cross section through glass bead (red) and dust sample (light blue)
	   at maximum intrusion for our 3D simulation (with $p_m = 0.26$~kPa,
	   top) and the drop experiment (bottom). The colour indicates
	   the spatially averaged filling factor. The density structure
	   underneath the glass bead match very well. Even the slight
	   compression along the tight intrusion channel can be reproduced.
	   In the simulated plot a gap between glass bead and the most dense
	   area is clearly visible. This is due to the smoothing of the
	   sphere and was discussed in Sect.\ \ref{sec:resolution-and-convergence}.
	   The gap has roughly the size of one smoothing length $h$. 
	   }
  \label{fig:fill-3Dfinal}
\end{figure}

(5) The comparison between the cross-sections through sphere and 
dust sample along the z-axis (Fig.\ \ref{fig:fill-3Dfinal}) reveals 
reasons for the excess of compressed volume. First, the
cross-section of the sphere is artificially enhanced by the smoothing
of its boundaries, which is inherent of the SPH method. One
effect of the smoothing is the existence of a gap between
sphere and dust sample, which was already discussed in Sect.\
\ref{sec:resolution-and-convergence} and is also clearly visible
in Fig.\ \ref{fig:fill-3Dfinal} (top). Hence, we assume
that the dust sample actually begins, where it has its maximum
compression. The fact that the sphere pokes out of the dust sample
a bit more than in the experiment has its reason also in the
artificial enlargement of the cross-section.
Second, it can be seen that in the experimental
reference (Fig.\ \ref{fig:fill-3Dfinal}, bottom) the compacted
region is much narrower and more concentrated underneath the
sphere. In the simulated result the compacted region is a bit
broader. This indicates that the shear strength seems to be
a bit smaller than we assume. Third, the compaction reaches
too high filling factors, which was already visible in the
cumulated volume diagram. 
Nevertheless, both cross-sections
match very well, especially with respect to the mediocre
resolution. Remarkably, even the slight intrusion channel on
the left and right side of the sphere, which features a slight
compression, can be reproduced.

\subsection{Bulk Modulus}
\label{sec:bulk-modulus}

\begin{figure*}
  \resizebox{\hsize}{!}{\includegraphics[angle=0]
			 {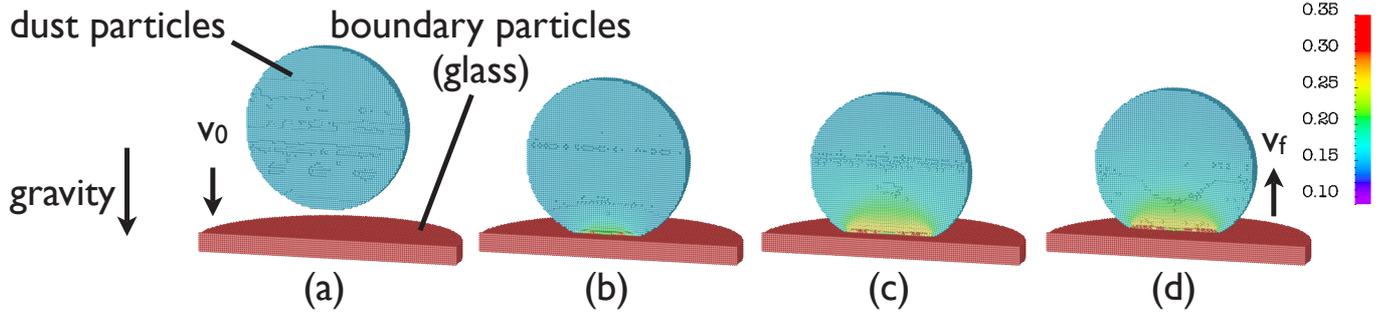}}
  \caption{Bouncing sequence for $t = 0$~ms (a), $t = 10$~ms (b), $t = 18$~ms (c), and
	   $t = 25$~ms (d). The colour code indicates the filling factor.
	   An aggregate consisting of dust particles (Sirono EOS,
	   diameter 1.0~mm) hits a solid surface simulated by boundary glass 
	   particles (Murnaghan EOS, diameter 1.6~mm, thickness 0.1~mm) with
	   a velocity of $v_0 = \, \rm 0.2~m\,s^{-1}$. For this simulation a bulk 
	   modulus of $K_0 = 5.0$~kPa and $p_m = 0.26$~kPa were used. The aggregate
	   hits the surface and starts to be compacted at its bottom (b). While
	   the plastic deformation at the bottom increases, the aggregate is also
	   deformed elastically: it gets broader (c). Eventually it leaves the
	   surface with a final velocity $v_f$ (d). It features a permanent compaction,
	   while the elastic deformation vanishes. (An animation of this figure is 
	   available in the online journal.)}
  \label{fig:bouncing-collection}
\end{figure*}

\begin{figure}
  \resizebox{\hsize}{!}{\includegraphics[angle=-90]
			 {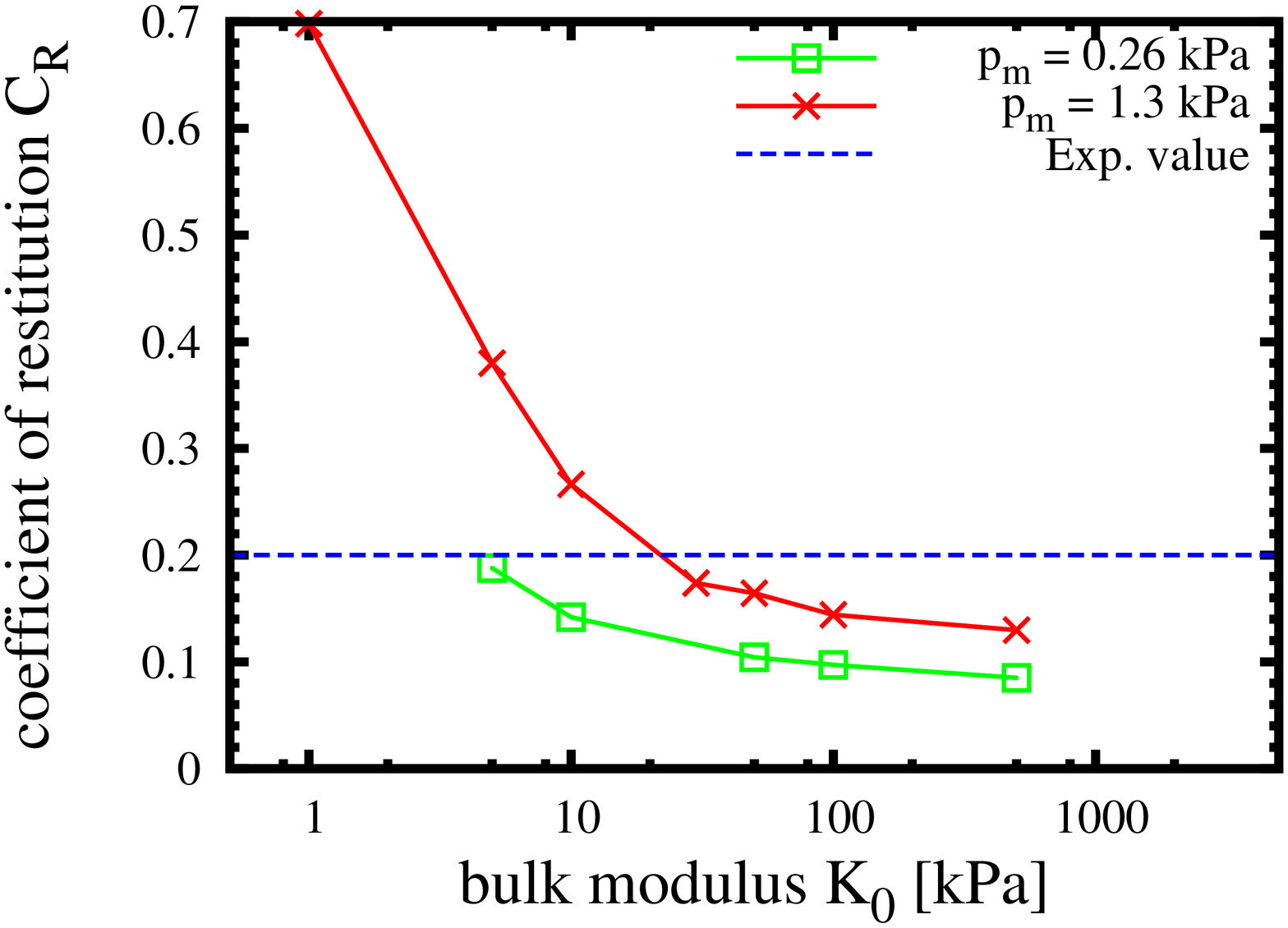}}
  \caption{Coefficient of restitution $\varepsilon_{\rm rest}$ over bulk modulus $K_0$ for values
	   of the mean pressure $p_m$ of the compressive strength relation. For
	   the lower $p_m = 0.26$~kPa (squares) more energy is dissipated by plastic
	   deformation. For this reason the coefficient of restitution
	   is significantly lowered compared to the higher $p_m = 1.3$~kPa (crosses).
	   Best agreement with the experimental reference of $\varepsilon_{\rm rest} = 0.2$ (dashed
	   line) is given for $K_0 \approx 5$~kPa and $K_0 \approx 20$~kPa, 
	   respectively.}
  \label{fig:bouncing-study}
\end{figure}

As it was mentioned in section \ref{sec:material_parameters} there are two estimates
for the bulk modulus $K_0$ for the uncompressed material with $\phi \approx 0.15$.
$K_0$ in our model is also the pre-factor for computing the bulk modulus for higher
filling factors with the aid of a power law (Eq.\ \ref{eq:bulk-modulus}). Since
these indirect findings, 1~kPa \citep[see][]{Weidling2009} and 300~kPa (from sound speed
measured by \citealt{BlumWurm2008,PaszunDominik2008}), differ by two orders of
magnitude, we try to find a suitable value for $K_0$ utilising a calibration experiment.

\begin{table}
\centering
\begin{tabular}{l c c l}
\hline\hline
Physical Quantity & Symbol & Value & Unit \\
\hline
\\
\textbf{Glass plate} & & \\
\hline
$\rm {Bulk~density}^{(*)}$ & $\rho_0$	& 2540 			& $\rm kg\,m^{-3}$ \\
$\rm {Bulk~modulus}^{(*)}$ & $K_0$ 	& $\rm 5 \times 10^9$ 	& Pa \\
$\rm {Murnaghan~exponent}^{(*)}$ & $n$		& 4			& - \\
Radius			&$r_{\rm plate}$& $0.8 \times 10^{-3}$  & m\\
Thickness		&$d_{\rm plate}$& $0.1 \times 10^{-3}$  & m \\
\\
\textbf{Dust sample} & & \\
\hline
Initial density		& $\rho_i$	& 300			& $\rm kg\,m^{-3}$ \\
Bulk density		& $\rho_s$	& 2000			& $\rm kg\,m^{-3}$ \\
Reference density	& $\rho_0'$	& 300			& $\rm kg\,m^{-3}$ \\
Initial filling factor	& $\phi_i$	& 0.15			& - \\
Bulk modulus		& $K_0$		& various		& Pa \\
ODC mean pressure	& $p_m$		& 260 and 1300		& Pa \\
ODC max.\ filling factor& $\phi_2$	& 0.58			& - \\
ODC min.\ filling factor& $\phi_1$	& 0.12			& - \\
ODC slope		& $\Delta$	& 0.58			& - \\
Impact velocity		& $v_0$		& 0.2			& $\rm m\,s^{-1}$ \\
Radius			& $r$		& $0.5 \times 10^{-3}$& $\rm m$ \\
\hline
\end{tabular}
\caption{Numerical parameters for the bouncing calibration setup. ODC stands for
omni-directional compression relation (Eq.\ \ref{eq:compressive-strength}). Quantities
marked by (*) represent the parameters for sandstone in \citet{Melosh1989}, which we
adopt for glass here.}
\label{tab:bouncing-parameters}
\end{table}

Simulating the low velocity collision setup by \citet{Weidling2009}, 
a 3D dust sphere ($\phi_i = 0.15$,1~mm diameter, 267737 particles, 
$l_c~=~12.5~{\mu}$m) drops onto 
a solid surface (cylindrical, 1.6~mm diameter, 0.1~mm thick, 115677 particles, 
$l_c~=~12.5~{\mu}$m) with initial velocity $v_0 = 0.2 \, \rm m\,s^{-1}$ 
(see Fig. \ref{fig:bouncing-collection}).
The material parameters are shown in Table \ref{tab:bouncing-parameters}. 
The bulk modulus $K_0$ is varied with respect to two values 
of the mean pressure $p_m$ in the compressive strength relation (0.26~kPa and 1.3~kPa).
During the impact a small region of the bottom of the dust sphere is compacted.
Then the deformed sphere bounces off the target with reduced velocity $v_f$ (see Fig.\ 
\ref{fig:bouncing-collection}). The latter effect was already observed in \citet{Guttler2009}
and demonstrates the ability of our code and the implemented porosity model to simulate
the elastic properties of the dust correctly. In this study we have doubled the resolution
and determine the coefficient of restitution $\varepsilon_{\rm rest} = v_f{v_0^{-1}}$ depending on $K_0$
(see Fig.\ \ref{fig:bouncing-study}).

The bouncing calibration setup is equivalent with two centrally
colliding dust aggregates at $0.4 \, \rm m\,s^{-1}$, thus, our results are comparable to
\citet{HeisselmannEtal:2007}: $\varepsilon_{\rm rest} \approx 0.2$ and $\approx 95 \, \%$ energy dissipation.

Based on the results of the previous section, where $p_m = 0.26$~kPa turned out to
be a good choice for the mean pressure, our results of this experiment favour
a bulk modulus $K_0 \approx 5$~kPa. This value is close to $K_0 = 1 \, \rm kPa$ 
computed by \citet{Weidling2009} with a rough model. Our simulations yield a
coefficient of restitution $\varepsilon_{\rm rest} = 0.19$ ($\approx 96$~\% energy dissipation)
for $K_0 = 5.0$~kPa, which is in excellent agreement with the experimental results.
On the other hand for $K_0 = 500$~kPa we find $\varepsilon_{\rm rest} = 0.09$ ($\approx 99$~\% energy
dissipation), which is too far away from the reference. A high value for the bulk
modulus $K_0$ as it is given by the sound speed measurements is therefore excluded.

Given the higher value $p_m = 1.3$~kPa for the compressive strength curve $\varepsilon_{\rm rest}$ is
raised for all choices of $K_0$. For $K_0 = 1.0$~kPa it yields $\varepsilon_{\rm rest} \approx 0.7$ 
and only $\approx 50 \, \%$ of the energy is dissipated. On the other hand for 
$K_0 = 300 \, \rm kPa$ we find $\varepsilon_{\rm rest} \approx 0.13$ equivalent to $\approx 98 \%$ 
energy dissipation. For $p_m = 1.3$~kPa it is harder to dissipate energy by 
plastic deformation. Hence, less energy is dissipated in the compaction process
and the coefficient of restitution is generally higher.

This bouncing experiment fixes our choice for the bulk modulus to $K_0 \approx 5$~kPa
while maintaining $p_m = 0.26$~kPa from the compaction experiment in the previous
section. The next section will show that this choice is also consistent with the
fragmentation behaviour of the dust aggregates.

\subsection{Fragmentation}
\label{sec:fragmentation}

Since the intended field of application of our calibrated SPH code will be the simulation
of pre-planetesimal collisions, it is of major importance to calibrate and test the
fragmentation behaviour of the simulated material. For this reason we simulate the
fragmentation experiment described already in the second part of Sect.\ 
\ref{sec:further-benchmark-experiments}.

In our simulation a dust aggregate ($\phi_i=0.35$, 189296 particles) hits a 
glass plate (188478 particles) from below with an impact velocity of 
$v_0 = 8.4 \, \rm m\,s^{-1}$. The spatial resolution 
($l_c = 8.0 \, \rm \mu m, h = 30.0 \, \rm \mu m$) was chosen such that 
a single particle has less than $5 \times 10^{-6}$ times the mass of
the whole aggregate ($6.8 \, \times \, 10^{-8} \, \rm kg$) to resolve the same fragment masses
as the experimental reference. Gravitation was taken into account. 
A list of the other relevant parameters can be found in 
Table \ref{tab:fragmentation-parameters}.

\begin{figure}
  \resizebox{\hsize}{!}{\includegraphics[angle=-90]
			 {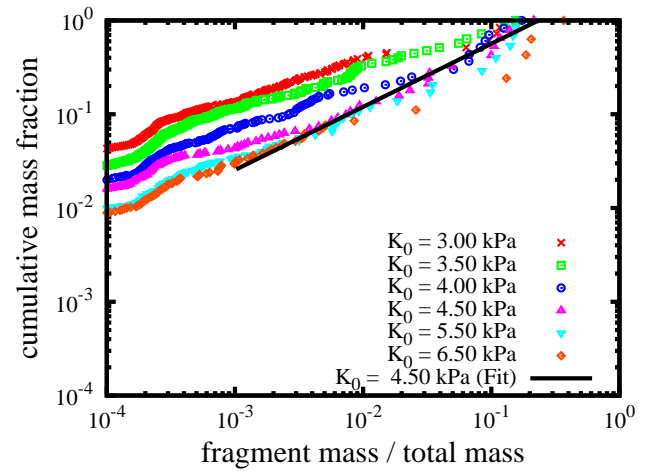}}
  \caption{Cumulative mass distribution of the fragments of a
	   dust aggregate impacting on a glass plate for different
	   values of $K_0$. For low fragment masses the 
	   shape of all simulated curves differs from the 
	   experimental curve in Fig.\ \ref{fig:frag_size_dist} 
	   due to the limited resolution
	   of the experimental setup.  An increase of $K_0$ leads to an
	   increase in the slope $\alpha$ of the power-law fit. The
	   best agreement with the experimentally measured slope 0.67
	   was found for the simulation with $K_0 = 4.5$~kPa.}
  \label{fig:fragdistribution}
\end{figure}

We investigated the influence of the bulk modulus on the fragmentation
behaviour. For this reason we varied $K_0$ from $3.0$~kPa to $6.5$~kPa.
For comparison with the reference experiment (Fig.\ \ref{fig:frag_size_dist})
we also plot our fragmentation data in a cumulative way
(Fig.\ \ref{fig:fragdistribution}) and find also a good a agreement with
the power-law distribution of Eq.\ \ref{eq:power-law-distribution}.
The simulation was evaluated after 0.8~ms simulated time. The mass of a
fragment is given by the sum of the mass of the SPH particles belonging
to it. We consider two fragments as separated when they are not linked
by SPH particles that interact with each other, i.e. when the closest
SPH particles of two fragments have a distance more than a smoothing
length.

\citet{Guttler2009} were facing the problem that using $K_0 = 300$~kPa
almost no fragmentation occurred (see Fig.\ 21 in this reference). Their
speculations about a modification of the shear strength in order to
resolve this problem proved to be wrong. The quantity with the most
impact on the fragmentation behaviour of the dust aggregate is its
bulk modulus. In general it can be said that an increase of the bulk
modulus leads to an increasing slope $\alpha$ of the fragment
distribution, whereas the size of the largest fragment (normalised
through the total mass of the projectile) $\mu$ roughly remains
constant at $\approx 20 \, \rm \%$ up to $K_0 = 6.0$~kPa. For
higher $K_0$ mainly small chunks and single SPH particles will burst
off the aggregate, which is only compacted but does not fragment.
This reproduces the situation described in \citet{Guttler2009}
and Sect.\ \ref{sec:experimental-reference}.

\begin{figure*}
  \resizebox{\hsize}{!}{\includegraphics[angle=0]
			 {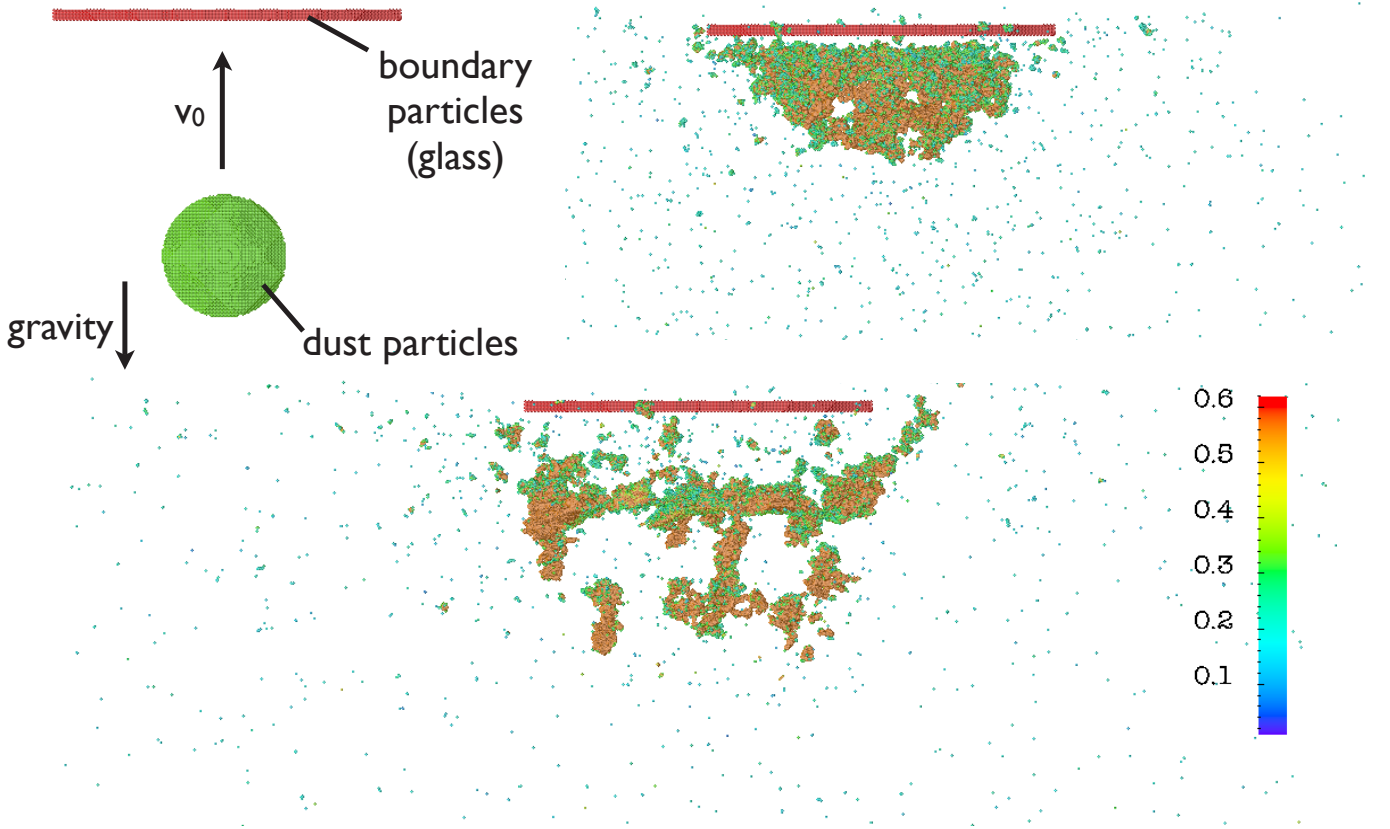}}
  \caption{Fragmentation sequence for $t = 0$~ms (top left), $t = 0.4$~ms (top right), and
	   $t = 0.8$~ms (bottom). The colour code indicates the filling factor.
	   A projectile consisting of dust particles (Sirono EOS,
	   diameter 0.57~mm) hits a solid surface simulated by boundary glass 
	   particles (Murnaghan EOS, diameter 1.6~mm, thickness 0.04~mm) with
	   an impact velocity of $v_0 = 8.4~m\,s^{-1}$. A bulk modulus of $K_0 = 4.5$~kPa
	   was used for this simulation. While many small pieces, often single SPH
	   particles, chip off, most of the aggregate is compacted to its maximum
	   filling factor $\phi_{\rm max} = 0.58$ and fractures. The bottom layer
	   of the impacting sphere is hardly compacted. This uncompacted layer is still
	   visible on the fragments.  (An animation of this figure is 
	   available in the online journal.)}
  \label{fig:fragmentation-collection}
\end{figure*}

We calibrate for $\alpha$, which is more sensitive to changes of
$K_0$ (see Table \ref{tab:fragmentation-results}). Given the
measured value of $\alpha = 0.67$ we find an excellent agreement with
our simulation using $K_0 = 4.5$~kPa, which yields $\alpha = 0.673 \, 
\pm \, 0.017$. Moreover this simulation reproduces also the
experimentally measured normalised mass of the largest fragment
$\mu = 0.22$ to a very high accuracy ($\mu = 0.234 \, \pm \, 0.007$).
The slight overestimation may be caused by the fact that the increase
in filling factor is not taken into account in the analysis of the
experimental data (see Sect.\ \ref{sec:experimental-reference}), whereas
in the simulation it is. The fragment distributions for different
$K_0$ and the best fit for the power-law are shown in Fig.\ 
\ref{fig:fragdistribution}. Setup and outcome of the
simulation are displayed in Fig. \ref{fig:fragmentation-collection}.
As already indicated in Sect.\ \ref{sec:bulk-modulus} the choice
of $p_m = 0.26$~kPa and $K_0 = 4.5$~kPa is consistent with the
results from the compaction and bouncing experiments. The
fragmentation experiment serves as a proof for the validity of
these choices and also for the consistency of the underlying 
porosity model. 

In contrast to the experiments we find no material sticking to
the glass plate. This is not possible due to the setup of
the simulation. As in Sect.\ \ref{sec:artificial-viscosity} we
use artificial viscosity to separate glass and dust materials.
This leads to an additional pressure on the dust material
which prevents sticking. 

\begin{table}
\centering
\begin{tabular}{l c c l}
\hline\hline
Physical Quantity & Symbol & Value & Unit \\
\hline
\\
\textbf{Glass plate} & & \\
\hline
$\rm {Bulk~density}^{(*)}$ & $\rho_0$	& 2540 			& $\rm kg\,m^{-3}$ \\
$\rm {Bulk~modulus}^{(*)}$ & $K_0$ 	& $\rm 5 \times 10^9$ 	& Pa \\
$\rm {Murnaghan~exponent}^{(*)}$ & $n$		& 4			& - \\
Radius			&$r_{\rm plate}$& $0.8 \times 10^{-3}$  & m\\
Thickness		&$d_{\rm plate}$& $0.04 \times 10^{-3}$  & m \\
\\
\textbf{Dust sample} & & \\
\hline
Initial density		& $\rho_i$	& 300			& $\rm kg\,m^{-3}$ \\
Bulk density		& $\rho_s$	& 2000			& $\rm kg\,m^{-3}$ \\
Reference density	& $\rho_0'$	& 700			& $\rm kg\,m^{-3}$ \\
Initial filling factor	& $\phi_i$	& 0.35			& - \\
Bulk modulus		& $K_0$		& various		& Pa \\
ODC mean pressure	& $p_m$		& 260			& Pa \\
ODC max.\ filling factor& $\phi_2$	& 0.58			& - \\
ODC min.\ filling factor& $\phi_1$	& 0.12			& - \\
ODC slope		& $\Delta$	& 0.58			& - \\
Impact velocity		& $v_0$		& 8.4			& $\rm m\,s^{-1}$ \\
Radius			& $r$		& $0.285 \times 10^{-3}$& $\rm m$ \\
\hline
\end{tabular}
\caption{Numerical parameters for the fragmentation calibration setup. ODC stands for
omni-directional compression relation (Eq.\ \ref{eq:compressive-strength}). Quantities
marked by (*) represent the parameters for sandstone in \citet{Melosh1989}, which we
adopt for glass here.}
\label{tab:fragmentation-parameters}
\end{table}

\begin{table}
\centering
\begin{tabular}{c c c}
\hline\hline
$K_0$ [kPa] & Slope $\alpha$ & Norm.\ largest fragment $\mu$ \\
\hline
\\
3.00	& 0.361	\, $\pm$ \, 0.004  & 0.200 \, $\pm$ \, 0.008 \\ 
3.50	& 0.429	\, $\pm$ \, 0.002  & 0.172 \, $\pm$ \, 0.002 \\ 
4.00	& 0.518	\, $\pm$ \, 0.011  & 0.230 \, $\pm$ \, 0.009 \\ 
4.25    & 0.523	\, $\pm$ \, 0.006  & 0.194 \, $\pm$ \, 0.004 \\
4.50	& 0.673	\, $\pm$ \, 0.017  & 0.234 \, $\pm$ \, 0.007 \\ 
4.75	& 0.834	\, $\pm$ \, 0.025  & 0.196 \, $\pm$ \, 0.005 \\ 
5.00	& 0.832	\, $\pm$ \, 0.063  & 0.198 \, $\pm$ \, 0.011 \\ 
5.50	& 0.836	\, $\pm$ \, 0.052  & 0.220 \, $\pm$ \, 0.010 \\ 
6.00	& 2.027	\, $\pm$ \, 0.121  & 0.171 \, $\pm$ \, 0.002 \\ 
6.50	& 0.910	\, $\pm$ \, 0.053  & 0.390 \, $\pm$ \, 0.013 \\ 
\hline
\end{tabular}
\caption{Results from the fragmentation calibration setup. The slope $\alpha$ of the
power-law is increasing with increasing bulk modulus $K_0$. Remarkably, the size of
the normalised biggest fragment remains nearly constant around $\mu \approx 0.2$ for
$K_0 \le 6.0$~kPa.}
\label{tab:fragmentation-results}
\end{table}

%______________________________________________________________

\section{Discussion}

In this work we presented a smooth particle hydrodynamics (SPH) code equipped
with extensions for continuum mechanics of solid bodies and an extended version
of the \citet{Sirono2004} porosity model. The code uses experimentally measured
macroscopic parameters such as tensile strength and a static compressive strength
relation. In \citet{Guttler2009} this code was used to determine a relation
for the shear strength and an estimate for the mean pressure $p_m$ for the
dynamic compressive strength relation (Eq.\ \ref{eq:compressive-strength}). 
The estimate was quite crude, though, due to the usage of 2D simulations only.

This work profoundly investigated the numerical properties of the SPH code.
Utilising the compaction calibration setup of \citet{Guttler2009} as an example
we determined an adequate size of the computational domain and adequate
numerical and spatial resolutions. We have shown that the results for this 
setup are converging for higher spatial resolutions. Boundary conditions are
necessary for all calibration setups presented in this work. Their treatment,
which is a difficult issue in SPH, was resolved by using boundary particles
with vanishing acceleration at every time step. The dissipative properties of the
artificial viscosity and its role for the stability of the simulation were
investigated. Artificial viscosity was used to separate dust and glass material which are highly
different in the ``stiffness'' of their equations of state.

We investigated the crucial differences between 2D and 3D compaction calibration
setup and proved the validity of the correction factor for the intrusion depth
(Eq. \ref{eq:correction-factor}) that was already used without confirmation
in \cite{Guttler2009}. A major difference between experiments and simulations,
a compaction that went too far down in the dust sample and along with this
a compaction of too much volume, was resolved by using the 3D setup.

Using a series of 2D simulations of the compaction calibration setup we predicted
a new, more accurate, value for the mean pressure $p_m$ of the dynamic
compressive strength relation. A 3D simulation with this value $p_m = 0.26$ kPa
was carried out. The results were compared with the experimental reference
using the same benchmark features as in \citet{Guttler2009}. We found a good
agreement.

In \citet{Guttler2009} it was already demonstrated that our code is in principle
capable of simulating the bouncing of dust aggregates. With the bouncing calibration 
setup we now investigated the ability of the SPH code of quantitatively simulating 
the elastic properties of the dust and the energy dissipation by compaction. We
found that the bulk modulus as well as the compressive strength relation have
significant influence. For smaller values of $p_m$ (low compressive strength)
and higher values of the bulk modulus more energy is dissipated. Using 
$p_m = 0.26$~kPa from the compaction calibration setup we deduced that the
uncompressed dust samples ($\phi \approx 0.15$) have a bulk modulus 
$K_0 \approx 5$~kPa. With this result we were able to decide between two differing
experimental values in favour of the indirect determination by \citet{Weidling2009}.

An important application of this code will be pre-planetesimal collisions. Therefore 
we also tested the code for its ability to simulate fragmentation of dust aggregates
quantitatively correct. For this we used a fragmentation calibration setup and
identified the bulk modulus as the quantity with the most influence on the
fragment distribution. Again using $p_m = 0.26$~kPa from the compaction
calibration setup we found the best agreement with the empirical reference
for the bulk modulus $K_0 = 4.5$~kPa. Remarkably this is consistent with
the findings from the bouncing calibration setup, which represents a test for
a totally different behaviour. The problem of almost no fragmentation
for $K_0 = 300$~kPa described in \citet{Guttler2009} was hereby traced back
to a wrong assumption for the bulk modulus.

\section{Conclusion}

\citet{Schafer2007} have shown that the outcome of pre-planetesimal collisions
crucially depend on their material properties. If the issue of collisional growth
is to be investigated by computer simulations, they suggest an implementation
of experimentally measured material parameters and thorough calibration and
comparison with laboratory experiments.

The SPH code presented in this work complies with their requirements.
Based on the experimental preparatory work by \citet{Guttler2009} this code
has been successfully tested with three kinds of calibration setups for the 
correct simulation of compaction, bouncing, and fragmentation.

We conclude that we have developed a tool that features a sufficient accuracy
for the investigation of the outcome of pre-planetesimal collisions in parameter
ranges that are not accessible to laboratory experiments. This is based on
the assumption that the macroscopic properties calibrated for in this work
do not change with increasing size. However, due to the thermodynamically simple
porosity model used in the SPH code the area of application is restricted
to a certain range of collision velocities. Collisions where shock propagation
cannot be neglected are outside the physically meaningful limit
of this model.
Changes of the mechanical properties with temperature and other thermodynamical
effects like sintering and melting cannot be simulated. An extension of
the porosity model and its implementation and calibration are necessary
to broaden the parameter range (e.g.\ supersonic impacts) and to consider 
a realistic environment for pre-planetesimal collisions in the protoplanetary
disc.

Nevertheless, within its area of application our code will be used to produce
a ``catalogue of collisions''. The influence of object sizes, porosity, relative
velocity, rotation, and impact parameter on the fragment distribution of 
pre-planetesimal collisions will be investigated in future work.

\begin{acknowledgements}

The authors want to thank Serena E.\ Arena for many fruitful discussions and helpful 
comments about the manuscript. 
The SPH simulations were performed on the university and bwGriD clusters of the
computing centre (ZDV) of the University of T\"ubingen. 
We thank M.-B.\ Kallenrode and the University of Osnabr\"uck for providing 
access to the XRT setup. 
We also thank the anonymous referee for clarifying comments and suggestions.
This project was funded
by the Deutsche Forschungsgemeinschaft within the Forschergruppe 759 ``The
Formation of Planets: The Critical First Growth Phase'' under grants Bl 298/7-1,
Bl 298/8-1, and Kl 650/8-1.
\end{acknowledgements}
%
%______________________________________________________________
%
\bibliographystyle{aa}
\bibliography{literature}

\end{document}